\documentclass[draftcls,onecolumn,11pt]{IEEEtran}

\usepackage{cite}      
\usepackage{graphicx}  
\usepackage{psfrag}    
\usepackage{subfigure} 
\usepackage{url}       
\usepackage{amssymb,amsmath,amscd,amsfonts,amstext,bbm, enumerate, ntheorem}
\usepackage{array}
\usepackage{multirow}

\usepackage[latin1]{inputenc}

\newcommand{\RR}{{\mathbb{R}}}
\newcommand{\NN}{{\mathbb{N}}}

\newcommand{\EE}{{\mathbb{E}}}
\newcommand{\T}{{\mathrm{T}}}
\newcommand{\tr}{{\mathrm{tr}}}
\newtheorem{lemma}{Lemma}
\newtheorem{corollary}{Corollary}
\newtheorem{proposition}{Proposition}
\newtheorem{theorem}{Theorem}

\def\adots{
  \mathinner{\mkern1mu\raise1pt\hbox{.}\mkern2mu\raise4pt\hbox{.}
  \mkern2mu\raise7pt\vbox{\kern7pt\hbox{.}}\mkern1mu}}
\def\build#1_#2^#3{\mathrel{
\mathop{\kern 0pt#1}\limits_{#2}^{#3}}}
\def\1{\mathbbm{1}}

\def\bx{\mathbf{x}}
\def\by{\mathbf{y}}

\theoremstyle{definition}

\newtheorem{assumption}{Assumption}

\newcommand{\cond}{{\, | \, }}

\def\Xset{\mathcal{X}}
\def\Eset{\mathrm{E}}
\def\rmd{\mathrm{d}}
\def\bw{\mathbf{w}}
\def\bv{\mathbf{v}}
\def\bp{\mathbf{p}}
\def\bq{\mathbf{q}}
\def\bx{\mathbf{x}}
\def\by{\mathbf{y}}
\def\rme{\mathrm{e}}
\def\rmi{\mathrm{i}}
\def\lip{\mathrm{Lip}}
\DeclareMathOperator*{\Cov}{Cov}

\IEEEoverridecommandlockouts

\begin{document}

\title{Error exponents for Neyman-Pearson detection of a continuous-time Gaussian Markov  process from noisy irregular samples}

\author{W. Hachem, E. Moulines and F. Roueff
\thanks{The authors are with Institut Telecom / Telecom ParisTech / CNRS LTCI, France.}
\thanks{e-mails:
\texttt{walid.hachem, eric.moulines, francois.roueff@telecom-paristech.fr}}}
\date{ }

\markboth{submitted to \emph{IEEE Trans. Information Theory}}
{\emph{et.al.}}

\maketitle

\begin{abstract}
This paper addresses the detection of a stochastic process in noise from irregular samples.
We consider two hypotheses. The \emph{noise only} hypothesis amounts to model the observations as a
sample of a i.i.d. Gaussian random variables (noise only). The  \emph{signal plus noise} hypothesis
models the observations as the samples of a continuous time stationary Gaussian process (the signal) taken at known but
random time-instants corrupted with an additive noise.
Two binary tests  are considered, depending on which assumptions is retained
as the null hypothesis.
Assuming that the signal is a linear combination of the solution of a multidimensional stochastic
differential equation (SDE), it is shown that the minimum Type II error probability decreases
exponentially in the number of samples when the False Alarm probability is
fixed. This behavior is described by \emph{error exponents} that are completely characterized.
It turns out that they are related with the asymptotic behavior of the Kalman Filter in
random stationary environment, which is studied in this paper. Finally, numerical illustrations of our claims are
provided in the context of sensor networks.
\end{abstract}


\newpage

\section{Introduction}
\label{sec-intro}
The detection of stochastic processes in noise has received a great deal of attention in the past decades, see for instance the
tutorial paper \cite{kai-poo-it98} and the references therein.
More recently, in the context of sensor networks, there has been a rising interest
in the analysis of detection performance  when the stochastic process
is sampled irregularly. An interesting approach in this direction has been initiated in \cite{mis-ton-sp08}
using \emph{error exponents} for assessing the performance of the optimal detection procedure.
In this paper, we follow this approach in the following general setting.
Given two integers $p$ and $q$, and $A$ a positive stable square matrix,
we consider the $q$-dimensional stochastic process defined as the stationary solution of the
stochastic differential equation (SDE)
\begin{equation}
\label{eq-diffusion-vector}
\rmd X(t) = - A \,  X(t) \, \rmd t + B \, \rmd W(t), \quad t \geq 0
\end{equation}
where $(W(t),\,t \geq 0)$ is a $p$-dimensional Brownian motion and where $B$
is a $q \times p$ matrix.
The SDE~\eqref{eq-diffusion-vector} is widely used to describe continuous time signals (see \cite{oks-livre03,
all-livre2007} and the references therein).
We are interested in the detection of the signal $(X(t),\,t\geq0)$ from a finite sample with missing data or irregular sample
spacing.
Let $C$ be a $d\times q$ matrix, $(T_n, n \geq 1)$ be a renewal sampling process and  $(V_n, n \geq 1)$ be a
sequence of i.i.d. Gaussian vectors with zero mean and identity covariance matrix.
We further assume that  $(X(t),\,t\geq0)$, $(V_n, n \geq 1)$ and $(T_n,\,n\geq1)$ are independent and that $A$, $B$ and $C$ are
known.
Based on the observed samples ${Y}_{1:N} = ( Y_1, \dots, Y_{N})$  and $T_{1:N} = ( T_1, \dots, T_{N})$, our goal is to decide
whether for $n=1,\dots, N$,  $Y_n = V_n$ or $Y_n = C X(T_n) + V_n$. The first situation will be referred to as the
\emph{noise hypothesis} and the second as the  \emph{signal plus noise hypothesis}.

The renewal hypothesis on $(T_n,\,n\geq 1)$ means that $T_n = \sum_{k=1}^n I_k$ where the $(I_k, k \geq 1)$ are nonnegative i.i.d. random variables
called \emph{holding times} with common distribution denoted by $\tau$.
This is a standard model for irregular sampling, see \cite{mic-01,mic-jor-mtns02, mis-ton-sp08}.
The most usual examples are:
\begin{itemize}
\item The Poisson point process. In this case, the $(I_k,\,k\geq1)$
are i.i.d. with exponential distribution $\tau(\rmd x) =
\lambda \exp(-\lambda x) \, \rmd x$.
This model has been considered in \cite{mic-01,mic-jor-mtns02,mis-ton-sp08}
to model the situation where the signal is measured in time by $N$
identical asynchronous sensors.
\item The Bernoulli process. This is the discrete time counterpart of the
Poisson process. In this case, the $(I_k,\,k\geq1)$
are i.i.d. and have geometric distribution up to a multiplicative time constant $S>0$, \textit{i.e.} $\tau(\{Sk\})= p
(1-p)^{k-1}$.
In practice, this model corresponds to a regular sampling with period $S$ for which observations are missing
at random, with failure probability $1-p$.
The regular sampling process corresponds to $p=1$.
\end{itemize}

Two binary tests are considered in this work: either {\sf H0} is the
noise hypothesis and {\sf H1} is the signal plus noise hypothesis, or the
opposite. Constraining the False Alarm
probability (probability for deciding {\sf H1} under {\sf H0}) to lie beneath
an $\varepsilon \in (0,1)$, it is well known that the minimum Type II error
probability is attained by the Neyman-Pearson test. It will be shown in this
paper that this minimum Type II error probability $\beta_N(\varepsilon)$
satisfies $\beta_N(\varepsilon) = \exp(-N ( \xi + o(1) ))$ as $N\to\infty$, where the error exponent $\xi$ does not depend on $\varepsilon$.
The error exponent $\xi$ is an indicator of the performance of the detection test. Its value will be shown to depend on the
distribution of signal (given by $A$, $B$ and $C$) and on the distribution of the sampling process (given by
$\tau$). An important goal in sensor network design is to optimize the sampling process.
Characterizing the error exponents offers useful guidelines in this direction. For instance,
\cite{sun-ton-poo-it06, mis-ton-sp08, cha-vee-sp03, cha-vee-it06, jay-sp07}
provide useful insights on such concrete problems as the choice of the
optimum mean sensor spacing possibly subject to a cost or a power
constraint. Other application examples are considered in
\cite{sun-mis-ton-eph-spmag06, sun-mis-ton-eph-jsac07}. In these contributions,
error exponents are used to propose optimum routing strategies
for conveying the sensors data to the fusion center.

In the context of Neyman-Pearson detection, these error exponents are
given by the limits of the likelihood ratios, provided that these limits exist.
Let ${\cal Z}_{1:N} = ( Z_1, \ldots, Z_{N} )$ be a
sequence of $N$ observed random vectors. Assume a binary test is
performed on this sequence, and assume that under hypothesis {\sf H0},
the distribution of ${\cal Z}_{1:N}$ has the density $f_{0,N}$, while under
{\sf H1}, this distribution has the density $f_{1,N}$.
Fix $\varepsilon \in (0,1)$ and let $\beta_N(\varepsilon)$ be the minimum
over all tests of the Type II error probability when the False Alarm
probability $\alpha$ is constrained to satisfy $\alpha \leq \varepsilon$.
Let
\[
{\cal L}_N({\cal Z}_{1:N}) =
\frac 1N \log\left(
\frac{f_{0,N}({\cal Z}_{1:N} )}{f_{1,N}({\cal Z}_{1:N} )}
\right)
\]
be the normalized Log Likelihood Ratio (LLR) associated with the received
${\cal Z}_{1:N}$. Then we have the following theorem
(see for instance \cite{che-it96} for a proof):
\begin{theorem}
\label{th-np}
Assume there is a real number $\xi$ such that the random variable
${\cal L}_N({\cal Z}_{1:N})$ satisfies
\begin{equation}
\label{eq-cvg-llr}
{\cal L}_N({\cal Z}_{1:N}(\omega)) \xrightarrow[N\to\infty]{} \xi
\quad \text{in probability under {\sf H0}} .
\end{equation}
Then for every $\varepsilon \in (0,1)$,
\[
- \frac 1N \log \beta_N(\varepsilon) \xrightarrow[N\to\infty]{} \xi \ .
\]
\end{theorem}
In the case where the $Z_i$ are i.i.d. under both hypotheses,
the analogue of
Theorem \ref{th-np} appeared in \cite{che-ams52} and is known as Stein's lemma.
The generalization to Theorem \ref{th-np} can be found in
\cite{lus-ruk-vaj-spa93, che-it96}.
In our case, the observed process is ${\cal Z}_{1:N} = (Z_1, \ldots, Z_{N})$
with $Z_n = ( Y_n, T_n )$, in other words, the measurements consist in the
sampled received signal and the sampling moments.
Let us consider that $Y_{1:N} = V_{1:N}$ under {\sf H0} and
$Y_{1:N} = (C X(T_n) + V_n)_{1\leq n \leq N}$ under {\sf H1}.
Recall that the probability distribution of ${T}_{1:N}$ does not depend on
the hypothesis to be tested. In these conditions, the LLR is given by
\begin{equation}
\label{eq-llr-conditional}
{\cal L}_N({\cal Z}_{1:N}) =
\frac 1N \log\left(
\frac{f_{0,N}({Y}_{1:N} \cond {T}_{1:N} )}{f_{1,N}({Y}_{1:N} \cond {T}_{1:N})}
\right)
\end{equation}
where $f_{0,N}( . \cond {T}_{1:N} )$ and $f_{1,N}( . \cond {T}_{1:N} )$ are
the densities of $Y_{1:n}$ conditionally to $T_{1:N}$ under {\sf H0} and
{\sf H1} respectively. It is clear that
$f_{0,N}( . \cond {T}_{1:N} ) = {\cal N}(0, 1)$. Being solution of the
SDE \eqref{eq-diffusion-vector}, the process $(X(t), \, t \geq 0)$ is a Gaussian process.
In consequence, $f_{1,N}( . \cond {T}_{1:N} ) = {\cal N}(0, R(T_{1:N}))$
where matrix $R(T_{1:N})$ is a covariance matrix that depends on $T_{1:N}$.
In the light of Theorem \ref{th-np} we need to establish
the convergence in probability of the Right Hand Side (RHS) of
Eq.~\eqref{eq-llr-conditional} towards a constant $\xi$, and to characterize
this constant, under the assumption $Y_{1:N} = V_{1:N}$. Alternatively,
if we consider that {\sf H0} is the Signal plus Noise hypothesis
$Y_{1:N} = (C X(T_n) + V_n)_{1\leq n \leq N}$, then we study the convergence
of $-{\cal L}_N$ under this assumption.

Theorem \ref{th-np} has been used for detection performance analysis in
\cite{sun-ton-poo-it06, ana-ton-swa-icassp07, sun-zha-ton-poo-sp08,
mis-ton-sp08}.
In the closely related Bayesian framework, error exponents have been
obtained in \cite{bha-it90, cha-vee-sp03, cha-vee-it06,jay-sp07}.
The closest contributions to this paper are
\cite{sun-ton-poo-it06, sun-zha-ton-poo-sp08, mis-ton-sp08} which
consider different covariance structure for the process and different sensors locations models.
In \cite{sun-ton-poo-it06}, Sung \emph{et.al.} consider the scalar version of
the SDE \eqref{eq-diffusion-vector} and a regular sampling. 
In \cite{sun-zha-ton-poo-sp08}, the authors
essentially generalize the results of \cite{sun-ton-poo-it06} to situations
where the sensor locations follow some deterministic periodic patterns.
In \cite{mis-ton-sp08}, the sampling process (sensor locations) is a renewal
process as in our paper, and the detector discriminates among two
scalar diffusion processes described by Eq.~\eqref{eq-diffusion-vector}.
Moreover, the observations are noiseless. 
Here, due to the presence of additive noise, our technique for establishing the existence of the error exponents
and for characterizing them differ substantially from \cite{mis-ton-sp08}.
We establish the convergence of the LLR
${\cal L}_N({\cal Z}_N)$ by studying the stability (ergodicity) of the Kalman filter,
using Markov chains techniques.

The paper is organized as follows.
In Section \ref{sec-main-results}, the main assumptions and notations are introduced
and the main results of the paper are stated. Proofs of these results are presented in Section
\ref{sec-proofs-vector}. A discussion of
the main results as well as some particular cases are presented in
Section \ref{sec-discussion}. Section \ref{sec-numerical-illustration}
is devoted to numerical illustrations.
The proofs in Section \ref{sec-proofs-vector} rely heavily on a theorem for
Markov chains stability shown in appendix \ref{anx-markov}. The other
appendices contain technical results needed in the proofs.

\section{The Error Exponents}
\label{sec-main-results}

We consider the following hypothesis test that we shall
call the ``H0-Noise'' test:
\begin{equation}
\label{eq-H0:noise-vector}
\begin{split}
{\sf H0} &: Y_n = V_n \quad \text{for} \ n=1, \ldots, N \\
{\sf H1} &: Y_n = C X(T_n) + V_n \quad \text{for} \ n=1, \ldots, N
\end{split}
\end{equation}
where $C$ is the $d \times q$ observation matrix and where
$(V_n)$ is an i.i.d. $d$-dimensional process with $V_1 \sim {\cal N}(0,1_d)$, where $1_d$ denotes the $d\times d$ identity matrix.
The assumptions are summarized below:
\begin{assumption}\label{ass:main}
The following assertions hold.
\begin{enumerate}[(i)]
\item The process $(X(t))_{t\geq 0}$ is a stationary solution of the stochastic differential
equation \eqref{eq-diffusion-vector} where $(W(t))_{t\geq 0}$ is a $p$-dimensional
Brownian motion.
\item $T_n = \sum_1^n I_k$ is a renewal process, that is, $(I_n)_{n\geq1}$ is a sequence of i.i.d. non-negative r.v.'s with
  distribution $\tau$ and $\tau(\{0\})<1$.
\item $(V_n)$ is a sequence of i.i.d. r.v.'s with $V_1 \sim {\cal N}(0, 1_d)$.
\item The processes $(X(t))_{t\geq0}$, $(T_n)_{n\geq1}$ and $(V_n)_{n\geq1}$ are independent.
\end{enumerate}
\end{assumption}

In order to be able to apply Theorem \ref{th-np}, we now develop the expression
of the LLR given by \eqref{eq-llr-conditional}. To that end, we derive the
expressions of the likelihood functions
$f_{0,N}({Y}_{1:N} \cond {T}_{1:N} )$ and $f_{1,N}({Y}_{1:N} \cond {T}_{1:N})$.
The density $f_{0,N}( . \cond {T}_{1:N})$ is simply the density
${\cal N}(0, 1_{Nd})$ of $(V_1, \ldots, V_N)$, therefore
\begin{equation}
\label{eq-f:noise}
f_{0,N}({Y}_{1:N} \cond {T}_{1:N} ) =
\frac{1}{\left(2 \pi \right)^{Nd/2}}
\exp\left( - \frac{1}{2} \sum_{n=1}^{N} Y_n^\T Y_n \right) \ .
\end{equation}
We now develop $f_{1,N}( . \cond {T}_{1:N})$ by mimicking the approach
developed in \cite{sch-it65} and in \cite{sun-ton-poo-it06}.
Solving Eq.~\eqref{eq-diffusion-vector} between $T_n$ and $T_{n+1}$,
the process $(X_n) = (X(T_n))_{n\geq 1}$ satisfies the recursion
\begin{equation}
\label{eq-difference-vector}
X_{n+1} = \rme^{-I_{n+1}A} X_n + U_{n+1}, \quad n \in \NN
\end{equation}
Let $Q(x)$ be the $q \times q$ symmetric nonnegative matrix defined by
\begin{equation}
\label{eq-def-Q(x)}
Q(x) = \int_0^{x} \rme^{- u A} B B^\T \rme^{- u A^\T} \rmd u \ .
\end{equation}
As $A$ is positive stable, the covariance matrix $Q(\infty)$
exists (by Lemma~\ref{lm-norm-A}) and is the unique solution of the so called
\emph{Lyapunov's equation}  $Q A^\T + A Q = B B^\T$
\cite[Chap.~2]{hor-joh-topics-91}.

Given the sequence $(I_n)$, the conditional distribution of the process
$(X_n)$ is characterized by this recursion equation and by the conditional
distribution of the sequence $(X_0,U_n)$, namely
it is a sequence of independent r.v.'s, $X_0 \sim {\cal N}(0, Q(\infty))$
and $U_n\sim{\cal N}\left(0, Q_n \right)$ where
$Q_n=Q(I_n)$ is the covariance matrix defined by~\eqref{eq-def-Q(x)},
see \cite[Chap.~5]{oks-livre03}. \\
Now we write
\begin{equation}
\label{eq-f:signal:produit}
f_{1,N}({Y}_{1:N} \cond {T}_{1:N}) =
\prod_{n=1}^{N}
f_{1,n,N}\left( {Y}_{n} \cond {Y}_{1:n-1}, {T}_{1:N}  \right)
\end{equation}
where $f_{1,n,N}(. \cond {Y}_{1:n-1}, {T}_{1:N} )$ is the density of $Y_n$
conditionally to $( {Y}_{1:n-1}, {T}_{1:N} )$.
In view of Eq.~\eqref{eq-difference-vector}, $Y_n = C X_n + V_n$ and the
assumptions on $(V_n)$, these conditional densities are Gaussian, in other
words
\begin{equation}
\label{eq-f1-conditional}
f_{1,n,N}\left( {Y}_{n} \cond  {Y}_{1:n-1}, {T}_{1:N}  \right)
=
\frac{1}{\det( 2 \pi \Delta_n)^{1/2}}
\exp\left( -\frac 12 (Y_n - \widehat{Y}_{n})^\T \Delta_n^{-1}
(Y_n - \widehat{Y}_{n}) \right)
\end{equation}
where $\widehat{Y}_{n} =
\EE\left[ Y_n \cond {Y}_{1:n-1}, {T}_{1:N} \right]$
and $\Delta_n = \Cov\left( Y_n - \widehat{Y}_{n} \cond {T}_{1:N} \right)$
are respectively the conditional expectation of the current observation $Y_n$ given the past observations and the
so-called innovation covariance matrix under {\sf H1}.
From Equations \eqref{eq-f:noise}, \eqref{eq-f:signal:produit} and
\eqref{eq-f1-conditional}, the LLR ${\cal L}_N$ writes
\begin{eqnarray}
{\cal L}_{N}({Y}_{1:N}, {T}_{1:N}) &=&
\frac{1}{N} \log f_{0,N}({Y}_{1:N} \cond {T}_{1:N})
-
\frac{1}{N} \log f_{1,N}( {Y}_{1:N} \cond {T}_{1:N})
\nonumber  \\
&=&
 \frac{1}{2N} \sum_{n=1}^{N} \log\det \Delta_n
+ \frac{1}{2N} \sum_{n=1}^{N} (Y_n - \widehat{Y}_{n})^T\Delta_n^{-1}
(Y_n - \widehat{Y}_{n})
-  \frac{1}{2N} \sum_{n=1}^{N}Y_n^\T Y_n \ .
\label{eq-expression-LLN-vector}
\end{eqnarray}
As $(Y_n)$ is described under {\sf H1} by the state equations
\begin{equation}
{\sf H1} :
\left\{ \begin{matrix} X_{n+1} &=& \rme^{-I_{n+1}A} X_n + U_{n+1} \\
Y_n &=& C X_n + V_n \end{matrix} \right. \quad \text{for} \ n=1, \ldots, N ,
\label{eq-Y:signal-vector}
\end{equation}
it is well known that $\widehat{Y}_n$ and $\Delta_n$ can be computed using the
Kalman filter recursive equations.
Define the $q\times 1$ vector $\widehat{X}_n$ and the $q\times q$ matrix $P_n$
as
\[
\widehat{X}_n = \EE[ X_{n} \cond  Y_{1:n-1}, {T}_{1:N} ] \quad
\text{and} \quad
P_{n} = \Cov\left( X_{n} - \widehat{X}_{n} \cond {T}_{1:N} \right) \ .
\]
The Kalman recursions which provide these quantities are
\cite[Prop. 12.2.2]{bro-dav-livre91}:
\begin{align}
\widehat{X}_{n+1} &= \rme^{-I_{n+1}A}
 \left( 1_q - P_n C^\T \left( C P_n C^\T + 1_d \right)^{-1} C \right)
\widehat{X}_{n}
+ \rme^{-I_{n+1}A} P_n C^\T \left( C P_n C^\T + 1_d \right)^{-1} Y_n
\label{eq-kalman-vector-X}  \\
P_{n+1} &= \rme^{-I_{n+1}A}
\left( 1_q - P_n C^\T \left( C P_n C^\T + 1_d \right)^{-1} C \right)
P_n \rme^{-I_{n+1}A^\T} + Q_{n+1} \;.
\label{eq-kalman-vector-P}
\end{align}
The recursion is started with the initial conditions $\widehat{X}_1 = 0$ and
$P_1 = Q(\infty)$. With these quantities at hand, $\widehat{Y}_{n}$ and
$\Delta_n$ are given by
\begin{equation}
  \label{eq:defDeltahatY}
  \widehat{Y}_{n} = C \widehat{X}_{n} \quad \text{and} \quad
\Delta_n = C P_n C^\T + 1_d \ .
\end{equation}
With these expressions at hand, our purpose is to study the asymptotic
behavior of ${\cal L}_{N}$ given by Eq.~\eqref{eq-expression-LLN-vector}
assuming that that $(Y_n)$ is i.i.d. with $Y_1 \sim {\cal N}(0, 1_d)$
(H0-Noise test).

In our analysis, we shall require Model \eqref{eq-diffusion-vector} to be
\emph{controllable},
\emph{i.e.}, $(A,B)$ satisfies
$\displaystyle{\sum_{\ell=0}^{q-1} A^\ell B B^\T {A^\T}^\ell > 0}$.
Recall that $(A,B)$ is controllable if and only if the matrix $Q(x)$ defined by Equation
\eqref{eq-def-Q(x)} is nonsingular for any $x > 0$
(see \cite[Chap.~6]{ros-livre70} for a proof).

The Kalman equations
\eqref{eq-kalman-vector-X}-\eqref{eq-kalman-vector-P} can be written
as a random iteration,
$$
W_{n} = F_{\eta_{n}}( W_{n-1}),\quad n\geq1\;,
$$
where $W_n = ( \widehat{X}_{n+1}, P_{n+1} )$, $\eta_n = ( I_{n+1}, Y_{n} )$ and, for any $\eta=(I,Y)\in\RR_+\times\RR^d$ and $\bw=(\bx,\bp)\in\RR^q
\times {\cal P}_q$ where ${\cal P}_q$ is the cone of $q \times q$ symmetric
nonnegative matrices,
\begin{equation}
  \label{eq:PsiEta}
F_\eta(\bw):=\left[
\begin{array}{cc}
\rme^{-IA}
 \left( 1_q - \bp C^\T \left( C \bp C^\T + 1_d \right)^{-1} C \right)
\bx+ \rme^{-I A} \bp C^\T \left( C \bp C^\T + 1_d \right)^{-1} Y \\
\rme^{-I A}
\left( 1_q - \bp C^\T \left( C \bp C^\T + 1_d \right)^{-1} C \right)
\bp \rme^{-I A^\T} + Q(I) \;.
\end{array}
\right]  \;.
\end{equation}
Under {\sf H0}, since $(\eta_n)$ is a sequence of i.i.d. random variables,
$(W_n)_{n\geq0}$ is a Markov chain starting at
$W_0=(0,Q(\infty))$.  Observe also that since the second component of $\Psi_\eta$ , denoted by $\tilde{F}_I(\bp)$ in the following,
does not depend on $x$, $(P_n)_{n\geq1}$ also is a Markov chain starting at
$P_1=Q(\infty)$ and since it neither depends on
$Y$, this is true under {\sf H1} as well.
Let $[0,Q(\infty)]$ denotes the subset of all matrices $\bp\in{\cal P}_q$ such that $\bp\leq Q(\infty)$.
It is easy to see that, for any $I\geq0$, $[0,Q(\infty)]$ is a stable set for $\tilde{F}_I$. Indeed, suppose that
$\bp\in[0,Q(\infty)]$, then
\begin{align*}
\tilde{F}_I(\bp)&=\rme^{-I A} \left( 1_q - \bp C^\T \left( C \bp C^\T + 1_d \right)^{-1} C \right)
\bp \rme^{-I A^\T} + Q(I)\\
&\leq \rme^{-I A} \bp \rme^{-I A^\T} + Q(I)\\
&\leq \rme^{-I A} Q(\infty) \rme^{-I A^\T} + Q(I) = Q(\infty)\;,
\end{align*}
by definition of $Q$ in~(\ref{eq-def-Q(x)}). Hence, in the following, we consider $(W_n)$ and $(P_n)$ as chains valued
in $\RR^q\times[0,Q(\infty)]$ and $[0,Q(\infty)]$, respectively.
We will denote by $\Pi$ and $\tilde{\Pi}$ the transition kernels associated to the chains $(W_n)$ (under {\sf H0}) and
$(P_n)$, respectively, that is, for test functions
$f$ and $\tilde{f}$ defined on $\RR^q\times [0,Q(\infty)]$ and $[0,Q(\infty)]$ ,
\begin{align*}
\Pi f(\bw)&=\EE[ f(F_\eta(\bw)) ],\quad \bw\in\RR^q\times [0,Q(\infty)]\\
\tilde{\Pi} \tilde{f}(p)&=\EE[ \tilde{f}(\tilde{F}_I(\bp)) ],\quad \bp\in[0,Q(\infty)]\;,
\end{align*}
where $\eta=(I,Y)$ is distributed according to the distribution
$\tau \otimes {\cal N}(0, 1_d)$.
We now state our main results.

\begin{proposition}
\label{prop-invariance-vector-P}
Suppose that Assumption~\ref{ass:main} holds with a state realization
$(A, B, C)$ such that $A$ is positive stable and $(A,B)$ is controllable.
Then the transition kernel $\tilde\Pi$ has a unique invariant distribution $\mu$.
\end{proposition}


The notation $| \bw |$ in the following proposition denotes some norm on $\RR^{q+q^2}$.
\begin{proposition}
\label{prop-invariance-vector-W}
Suppose that Assumption~\ref{ass:main} holds with a state realization
$(A, B, C)$ such that $A$ is positive stable and $(A,B)$ is controllable.
Then the transition kernel $\Pi$ has a unique invariant distribution $\nu$.
This distribution satisfies $\int | \bw |^r \rmd\nu( \bw) < \infty$ for any
$r > 0$.
Moreover, the distribution $\mu$ defined in Proposition~\ref{prop-invariance-vector-P} is the
marginal distribution $\mu(\cdot) = \nu(\RR^q \times \cdot)$.
\end{proposition}


The main result for the H0-Noise test in the vector case can now be stated.
\begin{theorem}
\label{th-H0:Noise-vector}
Suppose that Assumption~\ref{ass:main} holds with a state realization
$(A, B, C)$ such that $A$ is positive stable and $(A,B)$ is controllable.
Consider the hypothesis test \eqref{eq-H0:noise-vector}.
Let $\varepsilon \in (0,1)$. For a given $N$, let $\beta_N(\varepsilon)$
be the minimum of the Type II error probabilities over all tests for which
the false alarm probability $\alpha$ satisfies $\alpha \leq \varepsilon$.
Then, as $N\to\infty$, $N^{-1}\log \beta_N(\varepsilon) \to\xi_{\text{H0:Noise}}$, where
\begin{equation}
\label{eq-err-exp-H0:bruit-vector}
  \xi_{\text{H0:Noise}} =
\frac 12\int \left\{\log\det\left( C \bp C^\T + 1_d \right) + \tr\left[ C (\bx\bx^\T-\bp)  C^\T \left( C \bp C^\T + 1_d \right)^{-1} \right]\right\}
\; \rmd\nu(\bx, \bp) \in(0,\infty)\;,
\end{equation}
where the distribution $\nu$ is defined in Proposition~\ref{prop-invariance-vector-W}.
\end{theorem}


If we interchange the roles of {\sf H0} and {\sf H1} in
\eqref{eq-H0:noise-vector} (call this test the ``H0-Signal'' test), we obtain
the following result:
\begin{theorem}
\label{th-H0:Signal-vector}
Assume the setting of Theorem \ref{th-H0:Noise-vector} with the roles of
{\sf H0} and {\sf H1} interchanged.
Then, as $N\to\infty$, $N^{-1}\log \beta_N(\varepsilon) \to\xi_{\text{H0:Signal}}$, where
\begin{equation}
\xi_{\text{H0:Signal}} =
\frac 12 \left(
\tr \left( C Q(\infty) C^\T \right) -
\int \log\det\left( C \bp C^\T + 1_d \right) \rmd\mu(\bp)
\right)
  \in (0, \infty)  \;,
\label{eq-err-exp-H0:signal-vector}
\end{equation}
where the distribution $\mu$ is defined in Proposition~\ref{prop-invariance-vector-P}.
\end{theorem}

Note that the problem of existence and
uniqueness of $\mu$ as well as a study of its properties
in the case where the sampling process is a Bernoulli process
have been recently undertaken in \cite{kar-sin-mou-arxiv-09}.

\section{Proofs}
\label{sec-proofs-vector}


In this section, we prove Propositions~\ref{prop-invariance-vector-P}
and~\ref{prop-invariance-vector-W}, and
Theorems~\ref{th-H0:Noise-vector} and~\ref{th-H0:Signal-vector}.
All these results follow from an analysis of the Markov chains induced
by the transition kernels $\Pi$ and $\tilde{\Pi}$, or, equivalently, by the
random iteration functions $F_\eta$ and $\tilde{F}_I$ defined
in~(\ref{eq:PsiEta}).
We start with a series of preliminary results for which we will need the following notation and assumptions.
Assume that $(\eta,\eta_n)_{n\geq1}$ is a sequence of i.i.d. r.v.'s distributed according to the distribution
$\tau \otimes {\cal N}(0, 1_d)$, where $\tau$ is a distribution on $\RR_+$ such that $\tau(\{0\})<1$. We denote $\eta=(I,V)$
and $\eta_n=(I_n,V_n)$ for all $n\geq1$ in accordance with Assumption~\ref{ass:main}.
For any $\bx\in\RR^q$ and $\bp\in[0,Q(\infty)]$,  we define two Markov chains induced by $\Pi$ and $\tilde{\Pi}$
and starting at $\bw=(\bx,\bp)$ and $\bp$, respectively
\begin{equation*}
\left\{
\begin{split}
  &Z_0^{\bw}=\bw \quad\text{and}\quad  \tilde{Z}_0^\bp=\bp\\
  &Z_k^{\bw}=F_{\eta_k}(Z_{k-1}^{\bw})\quad\text{and}\quad \tilde{Z}_k^{\bp}=\tilde{F}_{I_k}(\tilde{Z}_{k-1}^{\bp}), \quad k\geq1\;.
\end{split}
\right.\end{equation*}
As noticed earlier, $\tilde{Z}_k^{\bp}$ corresponds to the second component of $Z_k^{\bw}$ for each $k$ and is valued in
$[0,Q(\infty)]$.
Finally we introduce the following notation for the Kalman gain matrix
$$
G(\bp)=\bp C^\T(1_d+C\bp C^\T)^{-1}\;,
$$
and the short-hand notation for $G(\tilde{Z}_k^{\bp})$ (the Kalman gain matrix at time $k$):
\begin{equation}
  \label{eq:KalmanGain}
  G_k^\bp=\tilde{Z}_k^{\bp}C^\T(1_d+C\tilde{Z}_k^{\bp}C^\T)^{-1},\quad k\geq 0\;,
\end{equation}
As for the Kalman transition matrix, we set
$$
\Theta(I,\bp)= \rme^{-I A}(1_q- G(\bp) C)\;,
$$
and the short-hand notation for $\Theta(I_k,G_k^{\bp})$ (the Kalman transition matrix at time $k$):
\begin{equation}
  \label{eq:KalmanTransition}
  \Theta_{k}^\bp=\rme^{-I_kA}(1_q-G_{k-1}^\bp C),\quad n\geq1\;.
\end{equation}
Using this notation and $Q_k=Q(I_k)$, the Kalman covariance update equation
$\tilde{Z}_k^{\bp}=\tilde{F}_{I_k}(\tilde{Z}_{k-1}^{\bp})$  can be expressed
for all $k\geq1$ as
\begin{align}
\tilde{Z}_k^{\bp}&= \Theta_{k}^{\bp} \tilde{Z}_{k-1}^{\bp}\rme^{-I_{k}A^\T} +
Q_k \nonumber \\
&= \Theta_{k} ^{\bp} \tilde{Z}_{k-1}^{\bp} \Theta_{k}^{\bp\,\T} +
\rme^{-I_{k}A} ( 1_q - G_{k-1}^{\bp} C )
\tilde{Z}_{k-1}^{\bp} C^\T G_{k-1}^{\bp\,\T} \rme^{-I_{k}A^\T} + Q_{k}
\label{eq-update-P-transition-matrix} \\
&= \Theta_{k}^{\bp}   \tilde{Z}_{k-1}^{\bp} \Theta_{k}^{\bp\,\T}
+ \overline{Q}_k\;,
\label{eq-relation-P-Q}
\end{align}
where
\begin{equation}
  \label{eq:Qbar}
\overline{Q}_k=
\rme^{-I_{k}A} G_{k-1}^{\bp}  G_{k-1}^{\bp\,\T} \rme^{-I_{k}A^\T}
+ Q_{k},\quad k\geq1\;.
\end{equation}
Finally we denote a product of successive Kalman transition matrices by
\begin{equation}
  \label{eq:KalmanTransitionProd}
 \Theta_{n,m}^\bp=\Theta_{n}^\bp\Theta_{n-1}^\bp\dots\Theta_{m+1}^\bp\;,\quad 0\leq m < n\;.
\end{equation}
Note that $\Theta_{n,n-1}=\Theta_n$. If $m=n$, we will use the convention $\Theta_{n,n}=1_q$.

We shall prove a moment contraction result on the sequence $(\Theta_{n,0}^\bp)_{n\geq1}$ (Lemma~\ref{lem:stabTheta})
and from this result and some algebra (mainly contained in Proposition~\ref{prop:main-Lipschitz-bounds}) deduce a moment
contraction condition on the random iteration functions $F_\eta$ and $\tilde{F}_I$. Then a general result on Markov chains
(Theorem~\ref{thm:basicMarkov} in the appendix), tailored for this kind of conditions, will allow to conclude the proofs of
our main results.

We first derive a deterministic bound for $\Theta_{n,m}^{\bp}$ based on~(\ref{eq-relation-P-Q}), which relies on a Lyapunov
function argument similar to that in \cite[Theorem~2.4]{guo:1994} and \cite[Sec. 4]{and-moo-siam81}.
In the following, we denote by $|x|$ the Euclidean norm of the vector $x$,
$\lambda_{\min}(H)$ and $\lambda_{\max}(H)$ the minimum and
maximum eigenvalue of the matrix $H$ and by $\|H\|$ its operator norm, $\|H\|=\lambda_{\max}(H^\T H)^{1/2}$.

\begin{lemma}
\label{lm-decrease-upsilon}
For any $0\leq m < n$, we have
\begin{equation}
\label{eq-bound-norm-upsilon}
\| \Theta_{n,m}^\bp \|^2 \leq
\| \tilde{Z}_n^{\bp} \| \, \| (\tilde{Z}_m^{\bp})^{-1} \| \,
 \prod_{k=m+1}^{n}
\left( 1 - \frac{\lambda_{\min}( Q_{k} )}{\| \tilde{Z}_k^{\bp} \|} \right) \ ,
\end{equation}
\end{lemma}


\begin{proof}
Obviously $\overline{Q}_{k} \geq Q_{k}$, hence
$\lambda_{\min}( \overline{Q}_{k} ) \geq \lambda_{\min}( Q_{k} )$.
Now, for a given $x_n \in \RR^q$, define the backward recursion
$x_k = \Theta_{k+1}^{\bp\,\T} x_{k+1}$ for $k$ decreasing from $n-1$ down to  $m$, and set
$V_k = x_k^\T \tilde{Z}_k^{\bp} x_k$ for $k=m, \ldots, n$. We have
\[
V_{n} - V_{n-1} =
x_{n}^\T \tilde{Z}_n^{\bp} x_{n} -
x_{n}^\T \Theta_{n}^{\bp} \tilde{Z}_{n-1}^{\bp} \Theta_{n}^{\bp\,\T} x_{n}
= x_{n}^\T \overline{Q}_{n} x_{n} \,
\]
by~(\ref{eq-relation-P-Q}), and moreover,
$x_{n}^\T \overline{Q}_{n} x_{n}
\geq | x_{n} |^2 \lambda_{\min}( \overline{Q}_{n} )
\geq | x_{n}|^2 \lambda_{\min}( Q_{n} )
\geq V_{n} \lambda_{\min}( Q_{n} ) / \|\tilde{Z}_n^{\bp}\| \ .
$
Hence,
$V_{n-1} \leq V_{n}
\left( 1 - \lambda_{\min}( Q_{n} ) / \|  \tilde{Z}_n^{\bp}\| \right)$.
Iterating, we obtain
\begin{equation}
\label{eq-V_n-V_m}
V_m \leq V_{n} \prod_{k=m+1}^{n}
\left( 1 - \frac{\lambda_{\min}( Q_{k} )}{\| \tilde{Z}_k^{\bp} \|} \right)
\leq | x_{n} |^2 \| \tilde{Z}_n^{\bp} \|
 \prod_{k=m+1}^{n}
 \left( 1 - \frac{\lambda_{\min}( Q_{k} )}{\| \tilde{Z}_k^{\bp}\|} \right) \ .
\end{equation}
On the other hand, by~(\ref{eq:KalmanTransitionProd}),
$V_m= x_m^\T \tilde{Z}_m^{\bp} x_m =
x_{n}^\T \Theta_{n,m}^{\bp} \tilde{Z}_m^{\bp} \Theta_{n,m}^{\bp\,\T} x_{n}$, hence
$\lambda_{\min}( \tilde{Z}_m^{\bp}) \ x_{n}^\T \Theta_{n,m}^{\bp} \Theta_{n,m}^{\bp\,\T} x_{n}
\leq V_m$.
This, with Inequality \eqref{eq-V_n-V_m}, implies \eqref{eq-bound-norm-upsilon}.
\end{proof}

\begin{lemma}\label{lem:stabTheta}
Assume that the matrix $A$ is positive stable and that the pair $(A,B)$ is controllable.
For any $r>0$, there exist $K>0$ and $\rho\in(0,1)$ such that
$$
\EE\left[\sup_{\bp\in[0,Q(\infty)]}\|\Theta_{n,m}^\bp\|^{2r}\right]\leq K\,\rho^{n-m},\quad 0\leq m < n\;.
$$
\end{lemma}


\begin{proof}
Recall that for $\bp\in[0,Q(\infty)]$, we have $\tilde{Z}^\bp_k\in[0,Q(\infty)]$ for all $k\geq1$.
Note that $G$ is continuous and, by Lemma~\ref{lm-norm-A},
$\sup_k\|\rme^{-I_kA}\|<\infty$. Hence,
$$
\tilde{Z}^*=\sup_{\bp\in[0,Q(\infty)]}\sup_{k\geq1}\|\tilde{Z}^\bp_k\|<\infty
\quad\text{and}\quad
\Theta^*=\sup_{\bp\in[0,Q(\infty)]}\sup_{k\geq1}\|\Theta_k^\bp\|<\infty\;.
$$
Let $0\leq m < n$. Let $\epsilon>0$ that we will chose arbitrarily small later. Denote $T=\inf\{k\geq m, I_k\geq\epsilon\}$.
Then we have, by~(\ref{eq-relation-P-Q}) and~(\ref{eq:Qbar}),
\begin{equation}
\label{eq:atT}
\lambda_{\min}(\tilde{Z}^\bp_T)\geq\lambda_{\min}(Q_T)\geq
\lambda_{\min}(Q(\epsilon))>0\;,
\end{equation}
by Lemma~\ref{lm-properties-Q}. We now write
\begin{align*}
\|\Theta_{n,m}^\bp\|^{2r}&=\sum_{k=m}^{n-1}
\|\Theta_{n,k}^\bp\Theta_{k,m}^\bp\|^{2r}\1_{T=k}+\|\Theta_{n,m}^\bp\|^{2r}\1_{T\geq n}\\
&\leq \sum_{k=m}^{n-1}\|\Theta_{n,k}^\bp\|^{2r}(\Theta^*)^{2r(k-m)}\1(T=k)+(\Theta^*)^{2r(n-m)}\1_{T\geq n}\;.
\end{align*}
For any $k<n$, applying Lemma~\ref{lm-decrease-upsilon} and the bound~(\ref{eq:atT}), we have, on the event $T=k$,
$$
\|\Theta_{n,k}^\bp\|^{2r}\leq (\tilde{Z}^* \lambda_{\min}(Q(\epsilon))^{-1})^{2r}\; \prod_{j=T+1}^{n}
\left( 1 - \frac{\lambda_{\min}( Q_{j} )}{\tilde{Z}^{*}} \right)^{2r} \ ,
$$
Observe that $T-m+1$ is a geometric r.v. with parameter $\tau_\epsilon=\tau([\epsilon,\infty))$ and that $(Q_{T+i})_{i\geq1}$
is i.i.d., independent of $T$, and follows the same distribution as $Q(I)$. Moreover, by  Lemma~\ref{lm-properties-Q},
$\lambda_{\min}( Q(I))>0$ for $I>0$, and since, $\tau(\{0\})<1$, we have
$\gamma=\EE[(1-\lambda_{\min}( Q(I) )/\tilde{Z}^{*})^{2r}]<1$. Thus we get
\begin{multline*}
\EE\left[\sup_{\bp\in[0,Q(\infty)]}\|\Theta_{n,m}^\bp\|^{2r}\right]\\
\leq (\tilde{Z}^* \lambda_{\min}(Q(\epsilon))^{-1})^{2r}\;\sum_{k=m}^{n-1}\gamma^{n-k}
(\Theta^*)^{2r(k-m)}\tau_\epsilon(1-\tau_\epsilon)^{k-m}+(\Theta^*)^{2r(n-m)}\sum_{k\geq
  n}\tau_\epsilon(1-\tau_\epsilon)^{k-m}\\
\leq \{(\tilde{Z}^* \lambda_{\min}(Q(\epsilon))^{-1})^{2r}\tau_\epsilon(n-m)+1\}\tilde{\rho}^{n-m}\;.
\end{multline*}
where we chose $\epsilon>0$ small enough so that $(\Theta^*)^{2r}(1-\tau_\epsilon)<1$ and set
$\tilde{\rho}=\gamma\vee\{(\Theta^*)^{2r}(1-\tau_\epsilon)\}<1$.
This gives the result for any $\rho\in(\tilde{\rho},1)$ by conveniently choosing $K$.
\end{proof}

\begin{proposition}
\label{prop:main-Lipschitz-bounds}
We have, for all $\bp,\bq\in{\cal P}_q$,
\begin{equation}
  \label{eq:LipschitzP1}
\tilde{Z}_n^{\bp}-\tilde{Z}_n^{\bq}=\Theta_{n,0}^\bp(\bp-\bq)\Theta_{n,0}^{\bq\T},\quad n\geq1\;.
\end{equation}
Moreover, there exists a constant $C>0$ such that, for all $\bp,\bq\in[0,Q(\infty)]$,
\begin{align}\label{eq:GnBound}
&  \|G_n^\bp-G_n^\bq\| \leq C \; \|\bp-\bq\|\; \|\Theta_{n,0}^\bp\| \; \|\Theta_{n,0}^\bq\|\;,&\quad n\geq1\;,\\
\label{eq:ThetaBound}
&  \|\Theta_{n,m}^\bp-\Theta_{n,m}^\bq\| \leq C \; \|\bp-\bq\| \;
  \sum_{j=m+1}^{n}\|\Theta_{n,j}^\bq\|\,\|\Theta_{j-1,m}^\bp\|\,\|\Theta_{j-1,1}^\bq\|\,\|\Theta_{j-1,1}^\bp\| \;,&\quad  0\leq m < n\;.
\end{align}
\end{proposition}


\begin{proof}
Let us prove~(\ref{eq:LipschitzP1}). By induction, it is sufficient to show that
\begin{equation}\label{eq:inductionLipP1}
\tilde{F}_I(\bp)-\tilde{F}_I(\bq)=
\Theta(I,\bp)(\bp-\bq)\Theta^{\T}(I,\bq) \; .
\end{equation}
By continuity of $\tilde{F}_I$ and $\Theta(I,\cdot)$, we may assume that $\bp$ and $\bq$ are invertible.
In this case, the matrix inversion lemma gives that
\begin{equation}\label{eq:inversionLemma}
(\bp-\bp C^\T(C\bp C+ 1_d)^{-1} C \bp) = (\bp^{-1}+C^\T C)^{-1}\;,
\end{equation}
and the same is true with $\bq$ replacing $\bp$. Hence
\begin{align*}
\tilde{F}_I(\bp)-\tilde{F}_I(\bq)&=\rme^{-IA}\left[(\bp^{-1}+C^\T C)^{-1}-(\bq^{-1}+C^\T C)^{-1}\right]\rme^{-IA^\T}\\
&=\rme^{-IA}(\bp^{-1}+C^\T C)^{-1}\bp^{-1}\left[\bp-\bq\right]\bq^{-1}(\bq^{-1}+C^\T C)^{-1}\rme^{-IA^\T}\;.
\end{align*}
Using again~(\ref{eq:inversionLemma}) and the definition of $\Theta$, we get~(\ref{eq:inductionLipP1}), which achieves the
proof of~(\ref{eq:LipschitzP1}).

We now prove~(\ref{eq:GnBound}). Observe that $G$ is continuously differentiable on the compact set $[0,Q(\infty)]$.
Hence $\|G(\bp)-G(\bq)\|\leq C\|\bp-\bq\|$ for some constant $C>0$. Thus, since $G_n^\bp=G(\tilde{Z}^\bp_n)$, the
bound~(\ref{eq:GnBound}) follows from~(\ref{eq:LipschitzP1}).

Finally we prove~(\ref{eq:ThetaBound}). We have, for all $0\leq m < n$ (recall the convention $\Theta_{n,n}=\Theta_{m,m}=1_q$),
$$
\Theta_{n,m}^\bp-\Theta_{n,m}^\bq=\sum_{j=m+1}^{n}\Theta_{n,j}^\bq(\Theta_j^\bp-\Theta_j^\bq)\Theta_{j-1,m}^\bp\;.
$$
On the other hand, $\Theta_j^\bp-\Theta_j^\bq=\rme^{-I_jA}(G_{j-1}^\bq-G_{j-1}^\bp)C$, and~(\ref{eq:ThetaBound}) thus follows
from~(\ref{eq:GnBound}).
\end{proof}

We can now prove Propositions \ref{prop-invariance-vector-P},  Theorem~\ref{th-H0:Signal-vector}, Proposition
\ref{prop-invariance-vector-W} and Theorem~\ref{th-H0:Noise-vector}, mainly as consequences of Theorem~\ref{thm:basicMarkov}.

\subsection*{Proof of Proposition \ref{prop-invariance-vector-P} and Theorem~\ref{th-H0:Signal-vector}.}
Using~(\ref{eq:LipschitzP1}) in Proposition~\ref{prop:main-Lipschitz-bounds}, Lemma~\ref{lem:stabTheta} and the Hölder
inequality, we obtain that, for any $q>0$ there exists $C>0$ and $\alpha\in(0,1)$ such that
\begin{equation}
  \label{eq:CondiFortildePi}
\EE\left[|\tilde{Z}_n^\bp-\tilde{Z}_n^\bq|^{q}\right]\leq C\alpha^n,\quad \bp,\bq\in[0,Q(\infty)],\;n\geq1\;.
\end{equation}
This corresponds to Condition~(i) in Theorem~\ref{thm:basicMarkov}. Condition~(ii) is trivially satisfied for any $s$ and
$r=1$ since here $\Xset=[0,Q(\infty)]$ is a compact state space. Hence Proposition \ref{prop-invariance-vector-P} follows from
Theorem~\ref{thm:basicMarkov}(a).

Next we prove Theorem~\ref{th-H0:Signal-vector}.
By Theorem~\ref{th-np}, it is sufficient to prove that
$-{\cal L}_{N}({Y}_{1:N}, {T}_{1:N})$ as expressed in~(\ref{eq-expression-LLN-vector}) converges to $\xi_{\text{H0:Signal}}$
in probability when $Y_n=C X_n+V_n$ for all $n\geq1$.
Since, for all $n\geq1$, $P_n=\tilde{Z}_{n-1}^{Q(\infty)}$ and $\log\det\Delta_n$ is a Lipschitz function of $P_n$, we have
by Theorem~\ref{thm:basicMarkov}(b) that
\begin{equation}
  \label{eq:DeltaPart}
  \frac1N\sum_{n=1}^N\log\det\Delta_n\xrightarrow[N\to\infty]{\text{a.s.}}\int \log\det(C\bp C^\T+1_d)\;\mu(\rmd\bp) \;.
\end{equation}
This is true independently of the definition of $(Y_n)$ and hence will also be used in the proof of Theorem~\ref{th-H0:Noise-vector}
In contrast the specific definition of $(Y_n)$ here implies that $\hat{Y_n}=\EE[Y_n\mid Y_{1:n-1},T_{1:N}]$ and
$\Delta_n=\Cov(Y_n-\hat{Y_n})$. Hence $((Y_n-\hat{Y_n})^\T\Delta_n^{-1}(Y_n-\hat{Y_n}))_{n\geq1}$ is a sequence of i.i.d.
${\cal N}(0,1)$ r.v.'s, which yields
$$
\frac1N\sum_{n=1}^N(Y_n-\hat{Y_n})^\T\Delta_n^{-1}(Y_n-\hat{Y_n})\xrightarrow[N\to\infty]{\text{a.s.}} d\;.
$$
On the other hand, in~(\ref{eq-expression-LLN-vector}) this limit cancels with
\begin{equation}
  \label{eq:Vnsquare}
  \frac1N\sum_{n=1}^N V_n^\T V_n\xrightarrow[N\to\infty]{\text{a.s.}} d\;,
\end{equation}
which appears in the last term of~(\ref{eq-expression-LLN-vector}) when developing $Y_n^\T Y_n= V_n^\T V_n+X_n^\T C^\T C
X_n+2X_n^\T C^\T V_n$. Hence it only remains to show that
\begin{align}\label{eq:limitX_n}
&\frac1N\sum_{n=1}^N X_n^\T C^\T C X_n\xrightarrow[N\to\infty]{\text{a.s.}}  \tr (C Q(\infty) C^\T) \;, \\
\label{eq:limitX_nV_n}
&\frac1N\sum_{n=1}^N X_n^\T C^\T V_n  \xrightarrow[N\to\infty]{\text{a.s.}}   0 \;.
\end{align}
To this end, recall that $(X_n)$ is a Markov chain, whose distribution is
defined by the recurrence equation~(\ref{eq-difference-vector}) and the initial condition $X_0\sim{\cal N}(0,Q(\infty))$.
We shall establish the ergodicity of this Markov chain by again applying Theorem~\ref{thm:basicMarkov}.
For any $x\in\RR^q$, we denote by $(X_n^x)$ the Markov chain defined with the same recurrence equation but with initial
condition $X_0=x$. Then we have, by iterating,
\begin{align*}
  X_n^x= \rme^{-\sum_{j=1}^nI_j A} \; x + \sum_{k=1}^n \rme^{-\sum_{j=k+1}^nI_j A} \; U_k\;, n\geq 1\;.
\end{align*}
with the convention $\sum_{j=n+1}^nI_j=0$. Recall that, given $I_n$, the conditional distribution of $U_n$ is ${\cal
  N}(0,Q_n)$ and $Q_n=Q(I_n)\in[0,Q(\infty)]$. Hence $\EE[|U_n|^s\mid I_n]$ is a bounded r.v. for any $s>0$.
By Lemma~\ref{lm-norm-A}, we have $\EE[\|\rme^{-\sum_{j=k+1}^nI_j A}\|^s]\leq K\,(\EE[\rme^{-asI_1}])^{n-k}$ for same $K,s>0$.
Hence, we obtain, for any $s>0$, for some constants $C>0$ and $\alpha\in(0,1)$, for all $x,y\in\RR^q$,
\begin{align*}
&  \EE[|X_n^x-X_n^y|^s]\leq C \alpha^n\; (1+|x|^s+|y|^s)\;,\\
&  \EE[|X_n^x|^s]\leq C(1+|x|^s)\;.
\end{align*}
These are conditions (i) and (ii) of Theorem~\ref{thm:basicMarkov} with $r=p=1$.
Moreover, $(X_n)$ has a constant marginal distribution, namely ${\cal N}(0,Q(\infty))$, so that the invariant distribution
$\mu$ of Theorem~\ref{thm:basicMarkov}(a) is necessary $\mu={\cal N}(0,Q(\infty))$.
Now, applying Theorem~\ref{thm:basicMarkov}(b) and Theorem~\ref{thm:basicMarkov}(c), we get~(\ref{eq:limitX_n})
and~(\ref{eq:limitX_nV_n}), with $a=2$ and $a=1$ respectively. \\
To achieve the proof of Theorem~\ref{th-H0:Signal-vector}, it remains to prove
that $\xi_{\text{H0:Signal}} > 0$. This results from
$\log\det(C{\bf p}C^\T + 1_d) < \tr( C Q(\infty) C^\T)$ for every
${\bf p} \in [0, Q(\infty)]$.

\subsection*{Proof of Proposition \ref{prop-invariance-vector-W} and  Theorem~\ref{th-H0:Noise-vector}.}
Let $\bw=(\bx,\bp)\in\RR^q\times[0,Q(\infty)]$. We denote the first component of $Z_k^\bw$ by $\underline{Z}_k^\bw$ so that
$Z_k^\bw=(\underline{Z}_k^\bw\,\,\tilde{Z}_k^\bp)$.
Using the notation introduced above, we have, for all $k\geq1$,
$\underline{Z}_k^\bw=\Theta_k^\bp \underline{Z}_{k-1}^\bw+\rme^{-I_kA}G_{k-1}^\bp Y_{k-1}$, and, by iterating,
$$
\underline{Z}_n^\bw=\Theta_{n,0}^\bp\bx+\sum_{k=1}^{n}\Theta_{n,k}^\bp\rme^{-I_kA}G_{k-1}^\bp Y_{k-1},\quad n\geq1\;.
$$
By continuity of $G$, it is bounded on the compact set $[0,Q(\infty)]$, hence $\sup_{\bp,n}\|G_k^\bp\|<\infty$.
Also by Lemma~\ref{lm-norm-A}, $\sup_k\|\rme^{-I_kA}\|<\infty$. Applying these bounds, Lemma~\ref{lem:stabTheta},
the Minkowski Inequality and the Hölder Inequality in the previous display, we obtain, for any $s>0$ and some constant
$C>0$,
\begin{equation}
  \label{eq:CondiiForPi}
\EE[|\underline{Z}_n^\bw|^s]\leq C \, (1+|\bx|^s),\quad\bw=(\bx,\bp)\in\RR^q\times[0,Q(\infty)],\,n\geq1\;.
\end{equation}
Since the second component of $Z_k^\bw$ stays in the compact set $[0,Q(\infty)]$, this implies
Condition~(i) in Theorem~\ref{thm:basicMarkov} with $r=1$ for the complete chain $(Z_k^\bp)_{k\geq0}$.

Let now $\bv=(\by,\bq)\in\RR^q\times[0,Q(\infty)]$. We have
$$
\underline{Z}_n^\bw-\underline{Z}_n^\bv=\Theta_{n,0}^\bp\bx-\Theta_{n,0}^\bq\by+
 \sum_{k=1}^{n}(\Theta_{n,k}^\bp-\Theta_{n,k}^\bq)\rme^{-I_kA}G_{k-1}^\bp Y_{k-1}
+ \sum_{k=1}^{n}\Theta_{n,k}^\bq\rme^{-I_kA} (G_{k-1}^\bp- G_{k-1}^\bq) Y_{k-1}\;.
$$
Note that, using Lemma~\ref{lem:stabTheta}, the bounds~(\ref{eq:GnBound}) and~(\ref{eq:ThetaBound}) in
Proposition~\ref{prop:main-Lipschitz-bounds}, the Hölder Inequality and the Minkowski Inequality, we obtain, for any $r>0$
and some constants $C>0$, and $\rho\in(0,1)$ not depending on $\bp,\bq$,
$$
\EE[|G_{n}^\bp-G_{n}^\bq|^r]\leq C\rho^n\quad\text{and}\quad \EE[|\Theta_{n,m}^\bp-\Theta_{n,m}^\bq|^r]\leq C\rho^n,\quad 0\leq m< n\;.
$$
Using these bounds, Lemma~\ref{lem:stabTheta} and the
previous two displays, we thus obtain, for any $q>0$ and some constants $C>0$,
and $\alpha\in(0,1)$ not depending on $\bw,\bv$,
$$
\EE\left[|\underline{Z}_n^\bw-\underline{Z}_n^\bv|^{q}\right]\leq C\alpha^n\,(1+|\bx|^q+|\by|^q),\quad n\geq1\;.
$$
This, with~(\ref{eq:CondiFortildePi}), implies  Condition~(i)
in Theorem~\ref{thm:basicMarkov} with $p=1$ for the chain $(Z_k^\bp)_{k\geq0}$.
Hence Theorem~\ref{thm:basicMarkov}(a) applies, which yields the conclusions of Proposition~\ref{prop-invariance-vector-W}.

We now prove Theorem~\ref{th-H0:Noise-vector}, that is,
by Theorem~\ref{th-np}, we prove that
${\cal L}_{N}({Y}_{1:N}, {T}_{1:N})$ converges to $\xi_{\text{H0:Noise}}$ when $Y_n=V_n$ for all $n\geq1$.
Some of the terms appearing in~(\ref{eq-expression-LLN-vector}) are identical to the case where $Y_n=C X_n+V_n$ for all
$n\geq1$ investigated for the proof of Theorem~\ref{th-H0:Signal-vector}.
Writing
$$
(Y_n-\hat{Y_n})^\T\Delta_n^{-1}(Y_n-\hat{Y_n})=V_n^\T\Delta_n^{-1}V_n+2V_n^\T\Delta_n^{-1}C\hat{X_n}+\hat{X}_n^\T
C^\T\Delta_n^{-1}C\hat{X}_n\;,
$$
and using $\mu=\nu(\RR^q,\cdot)$,~(\ref{eq:DeltaPart}),~(\ref{eq:Vnsquare}) and some algebra, it is in fact sufficient to
prove that
\begin{align}
\label{eq:limitX_nX_nnoise}
&\frac1N\sum_{n=1}^N \hat{X}_n^\T C^\T\Delta_n^{-1}C\hat{X}_n\xrightarrow[N\to\infty]{\text{a.s.}}
\int \bx^\T C^\T(C\bp C^\T+1_d)^{-1}C \bx\;\rmd\nu(\bx,\bp)\;,\\
\label{eq:limitV_nV_nnoise}
&\frac1N\sum_{n=1}^N V_n^\T\Delta_n^{-1}V_n\xrightarrow[N\to\infty]{\text{a.s.}}
\int \tr (C\bp C^\T+1_d)^{-1}\;\rmd\mu(\bp)\;,\\
\label{eq:limitX_nV_nnoise}
&\frac1N\sum_{n=1}^N  V_n^\T\Delta_n^{-1}C\hat{X_n}\xrightarrow[N\to\infty]{\text{a.s.}} 0\;.
\end{align}
Now, these limits hold by observing that $(\hat{X}_n,P_n)=Z_{n-1}^{(0,Q(\infty))}$
and by apply Theorem~\ref{thm:basicMarkov}(b) with $a=1$ for~(\ref{eq:limitX_nX_nnoise}),
Theorem~\ref{thm:basicMarkov}(c) with $a=1$ for~(\ref{eq:limitX_nV_nnoise}) and
Theorem~\ref{thm:basicMarkov}(c) with $a=2$ for~(\ref{eq:limitX_nV_nnoise}). \\
It remains to prove that $\xi_{\text{H0:Noise}} > 0$. From Equation
\eqref{eq-err-exp-H0:bruit-vector},
$\xi_{\text{H0:Noise}} \geq \int f({\bf p})
\rmd \mu({\bf p})$ where
$f({\bf p}) = 0.5 \left( \log\det( C {\bf p} C^\T + 1_d) -
C {\bf p} C^\T ( C {\bf p} C^\T + 1_d )^{-1} \right)$. This function satisfies
$f({\bf p}) \geq 0$ and $f({\bf p}) = 0$ if and only if $C {\bf p} C^\T = 0$.
Let $\tilde{Z} \in [0, Q(\infty)]$ be a random variable with the invariant
distribution $\mu$, and assume that $C \tilde{Z} C^\T = 0$ with probability
one. From Equation \eqref{eq:PsiEta} we have with probability one
\begin{multline*}
0 = C \tilde{F}_I(\tilde{Z}) C^\T =
C \rme^{-I A} \tilde{Z} \rme^{-I A^\T} C^\T -
C \rme^{-I A} \tilde{Z} C^\T \left( C \tilde{Z} C^\T + 1_d \right)^{-1} C
\tilde{Z} \rme^{-I A^\T} C^\T + C Q(I) C^\T \\
=
C \rme^{-I A} \tilde{Z} \rme^{-I A^\T} C^\T +
C Q(I) C^\T
=
C Q(I) C^\T
\end{multline*}
Due to the controllability of $(A,B)$ and the fact that $\tau(\{0\}) < 1$, this
is a contradiction. Therefore $f(\tilde{Z}) > 0$ with probability one, hence
$\xi_{\text{H0:Noise}} > 0$, which achieves the proof of
Theorem~\ref{th-H0:Noise-vector}.

\section{Particular Cases, Discussion}
\label{sec-discussion}

Different particular cases and limit situations will be considered in this
section. We begin with the case where the sampling is regular, \emph{i.e.},
$I_1$ is equal to a constant that we take equal to one without loss of
generality. We then consider the case where the holding times are large
with high probability, \emph{i.e.}, the sensors tend to be far apart. Finally,
we consider the particular case where the SDE \eqref{eq-diffusion-vector} is
a scalar equation. \\
All proofs are deferred to Appendix \ref{anx-proofs-particular}.

\subsection*{Regular sampling}
When the sampling is regular, the model for $(Y_n)$ under {\sf H1}
(see Eqs.~\eqref{eq-Y:signal-vector}) is a general model for stable Gaussian
multidimensional ARMA processes corrupted with a Gaussian white noise. In this
case we denote by $\Phi = \rme^{-I_{n+1}A} = \exp(-A)$ and by
$Q = Q(1) = \int_0^1 \exp(-uA) BB^\T \exp(-uA^\T)\, \rmd u$ the state transition
matrix and the excitation covariance matrix respectively.

\begin{proposition}[Regular sampling]
\label{prop-regular-sampling}
In the setting of Theorem \ref{th-H0:Noise-vector}, assume that $I_1 = 1$
with probability one. Then
$- N^{-1} \log \beta_N(\varepsilon) \xrightarrow[N\to\infty]{} \xi_{\text{H0:Noise}}$
with
\begin{multline}
\xi_{\text{H0:Noise}} =
\frac 12 \left( \log\det\left( C P_{\text{R}} C^\T + 1_d \right)
- \tr\left[ C P_{\text{R}} C^\T \left( C P_{\text{R}} C^\T + 1_d \right)^{-1}
     \right] \right. \\
\left. + \tr\left[ C \Sigma C^\T \left( C P_{\text{R}} C^\T + 1_d \right)^{-1}
\right] \right)
\label{eq-exposant-H0:Noise-regular}
\end{multline}
where $P_{\text{R}}$ is the unique solution of the matrix equation
\begin{equation}
\label{eq-riccati}
P = \Phi P \Phi^\T -
\Phi P C^\T \left( C P C^\T + 1_d \right)^{-1} C P \Phi^\T
+ Q \,
\end{equation}
and where the $q\times q$ symmetric matrix $\Sigma$ is the unique solution of
the matrix linear equation
\begin{equation}
\label{eq-Sigma-regular-sampling}
\Sigma - \Phi (1_q-GC) \Sigma (1_q-GC)^\T \Phi^\T = \Phi G G^\T \Phi^\T
\end{equation}
with $G = P_{\text{R}} C^\T \left( C P_{\text{R}} C^\T + 1_d \right)^{-1}$. \\
Furthermore, when the roles of {\sf H0} and {\sf H1} are exchanged
(Theorem \ref{th-H0:Signal-vector}), then
\begin{equation}
- \frac 1N \log \beta_N(\varepsilon)
\xrightarrow[N\to\infty]{}
\xi_{\text{H0:Signal}} =
\frac 12 \left(
\tr \left( C Q(\infty) C^\T \right) -
\log\det\left( C P_{\text{R}} C^\T + 1_d \right)
\right) \ .
\label{eq-exposant-H0:Signal-regular}
\end{equation}
\end{proposition}

\vspace*{0.03\columnwidth}

Equation \eqref{eq-riccati} is the celebrated discrete algebraic Riccati
equation. Its solution $P_{\text{R}}$ is the asymptotic (steady state)
error covariance matrix when the sampling is regular. The matrix
$G = P_{\text{R}} C^\T \left( C P_{\text{R}} C^\T + I \right)^{-1}$ is the
Kalman filter steady state gain matrix \cite[Chap.~4]{and-moo-livre79}.

\subsection*{Large Holding Times}

We now study the behavior of the error exponents when the holding times
are large with high probability. We shall say that a family $(\tau_s)$ of
probability distributions on $\RR_+$ ``escapes to infinity'' as $s\to\infty$ if
\[ \forall K > 0, \, \tau_s([0,K]) \xrightarrow[s\to\infty]{} 0 . \]
In order to study the large holding time behavior of the error exponents,
we index the distribution of the holding times by $s$ and assume that
$\tau_s$ escapes to infinity.
A typical particular case that illustrates this situation is when we assume
that the $I_n$ are equal in distribution to $s \bar{I}$ where $\bar{I}$ is
some nonnegative random variable, and when we study the behavior of the error
exponents for large values of $s$.
\begin{proposition}[Large holding times]
\label{prop-large-s}
Assume $(\tau_s)$ escapes to infinity. The following facts hold true:
\begin{align}
\xi_{\text{H0:Noise}}
&\xrightarrow[s\to\infty]{}
\frac{1}{2} \left( \log\det\left(C Q(\infty) C^\T + 1_d\right) -
\tr\left[ C Q(\infty) C^\T \left(C Q(\infty) C^\T + 1_d\right)^{-1} \right]
\right) ,
\label{eq-xi-H0:Noise-large-s} \\
\xi_{\text{H0:Signal}} &\xrightarrow[s\to\infty]{}
\frac 12 \left( \tr\left( C Q(\infty) C^\T \right)
- \log\det\left(C Q(\infty) C^\T + 1_d\right)  \right) .
\label{eq-xi-H0:Signal-large-s}
\end{align}
\end{proposition}

\vspace*{0.03\columnwidth}

Given an $\RR^d$-valued i.i.d. sequence $(Y_n)$ such that
$Y_1 \sim {\cal N}(0, 1_d)$ under {\sf H0} and
$Y_1 \sim {\cal N}(0, C Q(\infty) C^\T + 1_d)$ under {\sf H1}, Stein's lemma
says that the associated Type II error exponent coincides with the
Kullback-Leibler divergence $D\left( {\cal N}(0, 1_d) \, \| \,
{\cal N}(0, C Q(\infty) C^\T + 1_d) \right)
= \text{RHS} \ \eqref{eq-xi-H0:Noise-large-s}$ while if
the roles of {\sf H0} and {\sf H1} are interchanged, the Type II error
exponent is
$D\left( {\cal N}(0, C Q(\infty) C^\T + 1_d)  \, \| \, {\cal N}(0, 1_d) \right)
= \text{RHS} \ \eqref{eq-xi-H0:Signal-large-s}$.
These results are expected: when $\tau_s$ escapes to infinity,
two consecutive samples $X(T_n)$ and $X(T_{n+1})$ will tend to be decorrelated,
and it will be realistic to approximate the process $(Y_n)$ received under
the signal hypothesis with an i.i.d. process which samples are distributed as
${\cal N}(0, C Q(\infty) C^\T + 1_d)$.

\subsection*{The Scalar Case}

In the scalar case, the SDE \eqref{eq-diffusion-vector} describes a so called
Ornstein-Uhlenbeck process
\begin{equation}
\label{eq-OH}
\rmd X(t) =  - a \, X(t)\, \rmd t + b \, \rmd W(t), \quad t \geq 0
\end{equation}
where $W(t)$ is a scalar Brownian motion and $(a,b)$ are known real non
zero constants. In our situation, $a > 0$ and the initial value $X(0)$
is independent from $W(t)$ and
follows the law ${\cal N}(0, Q(\infty))$ where the variance $Q(\infty)$ is
given by $Q(\infty) = b^2 / (2a)$. We
observe $(Y_n, T_n)_{1\leq n \leq N}$ where $(Y_n)$ is a scalar process and
we write the H0-Noise test as
\begin{align}
{\sf H0} &: Y_n = V_n \quad \text{for} \ n=1, \ldots, N \\
{\sf H1} &: Y_n = X(T_n) + V_n \quad \text{for} \ n=1, \ldots, N
\label{eq-H1-OH}
\end{align}
where the observation noise process $(V_n)$ is i.i.d. with $V_1 \sim {\cal N}(0,1)$.
Solving Equation \eqref{eq-OH} between $T_n$ and $T_{n+1}$ we obtain that
$X_n = X(T_n)$ is given by
\[
X_{n+1} = e^{-a I_{n+1}} X_n + U_{n+1}, \quad n \in \NN
\]
where $U_n \sim {\cal N}\left(0, Q_n = Q(\infty)(1 - e^{-2a I_{n}}) \right)$.
Statistically, the process $(Y_n)$ is completely described under {\sf H1} by the
scalars $a$ and $Q(\infty)$ and by the distribution $\tau$ of $I_1$, and so
are the error exponents. \\
The parameter $a$ controls the correlation strength between to samples
of $X(t)$ separated by a given time lag (the ``memory'' of the
Ornstein-Uhlenbeck process). If $I_1$ is integrable, we can assume
$\EE[I_1] = 1$ and include the mean holding time into $a$.
Turning to $Q(\infty)$, as $X(t) \sim {\cal N}(0, Q(\infty))$ for every
$t \geq 0$, we notice from \eqref{eq-H1-OH} that $Q(\infty)$ is simply equal
to the Signal to Noise Ratio $\text{SNR} = \EE[X_n^2] / \EE[V_n^2]$. \\

We begin by providing the error exponents expressions when the sampling
is regular. In the scalar case, it is easy to solve Equations
\eqref{eq-riccati} and \eqref{eq-Sigma-regular-sampling} in the statement of
Proposition \ref{prop-regular-sampling} and to obtain the error exponents
in closed forms:
\begin{corollary}[Corollary to Proposition \ref{prop-regular-sampling},
Regular sampling in the scalar case]
\label{cor-regular-sampling-scalar}
In the scalar case, assume $I_n = 1$ with probability one.
Put ${\Phi} = \exp(-a)$. Then in the setting of Theorem
\ref{th-H0:Noise-vector} the following holds true:
\begin{equation}
\label{eq-cvg-regular-test2}
- \frac 1N \log \beta_N(\varepsilon) \xrightarrow[N\to\infty]{}
\xi_{\text{H0:Noise}} =
\frac{1}{2}\left(
\log\left( 1 + P_{\text{R}} \right)
-
\frac{P_{\text{R}}}{P_{\text{R}}+1}
\left(
\frac{\Phi^2 P_{\text{R}}}
{P_{\text{R}}^2 + 2 P_{\text{R}} + 1 - \Phi^2}
- 1
\right)
\right)
\end{equation}
where
\[
P_{\text{R}} =
\frac{
(\text{SNR}-1)(1-{\Phi}^2)
+
\sqrt{ (\text{SNR}-1)^2(1-{\Phi}^2)^2 + 4 \text{SNR} (1-{\Phi}^2) }
}
{2} \ .
\]
In the setting of Theorem \ref{th-H0:Signal-vector}, we have
\[
\label{eq-cvg-refular-test1}
- \frac 1N \log \beta_N(\varepsilon) \xrightarrow[N\to\infty]{}
\xi_{\text{H0:Signal}} =
\frac{1}{2}\left( \text{SNR} - \log\left( 1 + P_{\text{R}} \right) \right)
\]
\end{corollary}

\vspace*{0.03\columnwidth}

The proof of this corollary is omitted.
We note that the result \eqref{eq-cvg-regular-test2} coincides with the one
stated in \cite[Theorem 1]{sun-ton-poo-it06}. \\

We now get back to a general distribution for the holding times and consider
the behavior of $\xi_{\text{H0:Signal}}$ with respect to $a$ and with
respect to the Signal to Noise Ratio:
\begin{proposition}
\label{prop-H0:signal-errexp-monotone}
Assume the scalar case.
In the setting of Theorem \ref{th-H0:Signal-vector}, the error exponent
$\xi_{\text{H0:Signal}}$ decreases as $a$ increases and
$\text{SNR} = Q(\infty)$ is
fixed, and $\lim_{a \to 0} \xi_{\text{H0:Signal}} = Q(\infty) / 2$.
Moreover, $\xi_{\text{H0:Signal}}$ increases as $Q(\infty)$ increases and $a$
is fixed.
\end{proposition}

\vspace*{0.03\columnwidth}

One practical implication of this proposition is the following: from the stand
point of the error exponent theory, when {\sf H0} stands for the presence of a
noisy O-U signal, one has an interest in choosing close sensors if one wants
to reduce the Type II error probability. This probability is reduced by
exploiting the correlations between the $X_n$. \\

In the setting of Theorem \ref{th-H0:Noise-vector}, the behavior of
$\xi_{\text{H0:Noise}}$ with respect to $a$ has been analyzed in the
regular sampling case only (Corollary \ref{cor-regular-sampling-scalar})
in \cite{sun-ton-poo-it06}. The authors of \cite{sun-ton-poo-it06} proved
that when $\text{SNR} \geq 0 \ \text{dB}$, $\xi_{\text{H0:Noise}}$ increases
with $a$ while when $\text{SNR} < 0 \ \text{dB}$, $\xi_{\text{H0:Noise}}$
admits a maximum with respect to $a$. By a numerical estimation of
$\xi_{\text{H0:Noise}}$ (see below), we observe a similar behavior in the
case of a Poisson sampling.
However, a more formal characterization of the behavior of
$\xi_{\text{H0:Noise}}$ for a general distribution $\tau$ seems to be
difficult.

\section{Numerical Illustration}
\label{sec-numerical-illustration}

Let $(\widehat{X}_\infty, P_\infty)$ be a random element of
$\RR^q \times [0, Q(\infty)]$ with distribution the invariant distribution
$\nu$ of the Markov process $(\widehat{X}_{n}, P_n)$. Then the error exponent
provided by Theorem \ref{th-H0:Noise-vector} can be also written
\[
\xi_{\text{H0:Noise}} = \frac 12 \EE \left[
\log\det\left( C {P}_\infty C^\T + 1_d  \right)
+
C \left( \widehat{X}_\infty \widehat{X}_\infty^\T - {P}_\infty\right) C^\T
\left(C {P}_\infty C^\T + 1_d\right)^{-1} \right]
\]
and the error exponent provided by Theorem \ref{th-H0:Signal-vector} is
$$
\xi_{\text{H0:Signal}} = \frac 12 \left( \tr\left( C Q(\infty) C^\T \right) -
\EE\left[ \log\det\left( C {P}_\infty C^\T + 1_d \right) \right] \right) .
$$
By the stability of the Markov chain $(\widehat{X}_{n}, P_n)$ shown
in Section \ref{sec-proofs-vector}, we estimate the error exponents by
simulating the Kalman Equations
\eqref{eq-kalman-vector-X}-\eqref{eq-kalman-vector-P} with $(Y_n)$ i.i.d.,
$Y_1 \sim {\cal N}(0,1_d)$, and by replacing the expectation operators in
the equations above with empirical means taken on
$(\widehat{X}_n, P_n)_{n=1,\ldots, N}$ for $N$ large enough. \\

\begin{figure}[t]
  \begin{center}
    \includegraphics[width=0.7\linewidth]{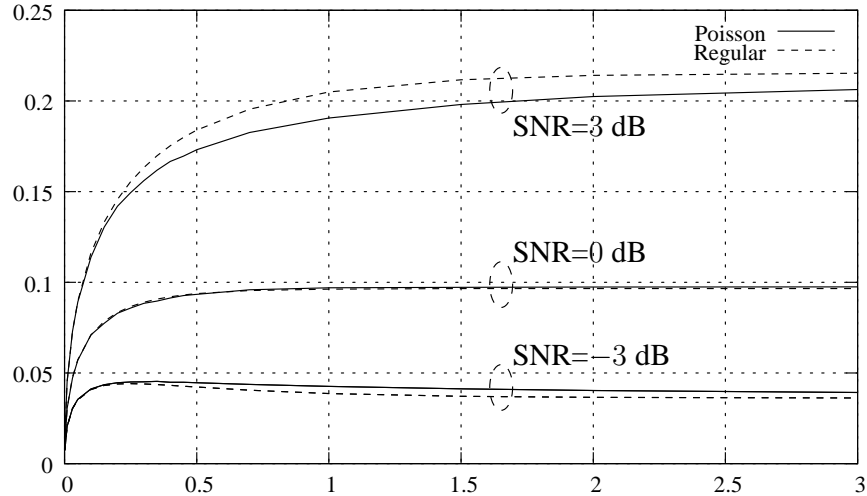}
  \end{center}
  \caption{Scalar case: $\xi_{\text{H0:Noise}}$ vs $a$ for
           $\text{SNR}=-3, 0$ and $3$ dB}
  \label{fig:H0-Noise-scalar}
\end{figure}

\begin{figure}[t]
  \begin{center}
    \includegraphics[width=0.7\linewidth]{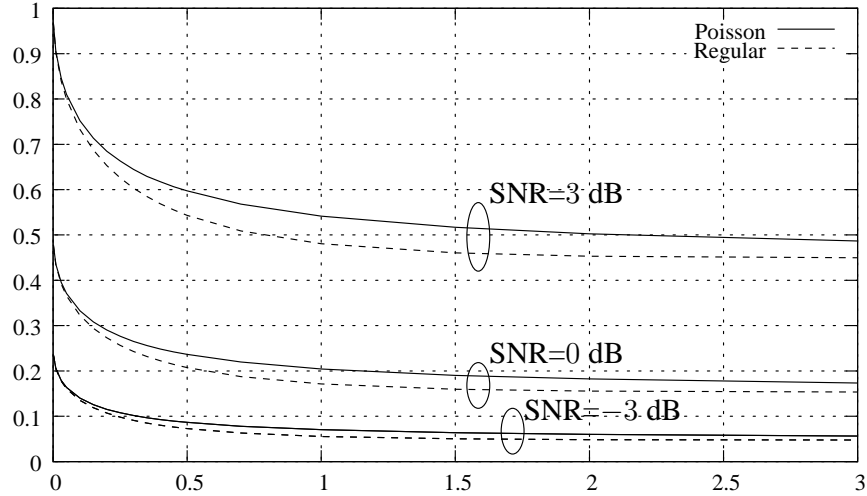}
  \end{center}
  \caption{Scalar case: $\xi_{\text{H0:Signal}}$ vs $a$ for
           $\text{SNR}=-3, 0$ and $3$ dB}
  \label{fig:H0-Signal-scalar}
\end{figure}

\begin{figure}[t]
  \begin{center}
    \includegraphics[width=0.7\linewidth]{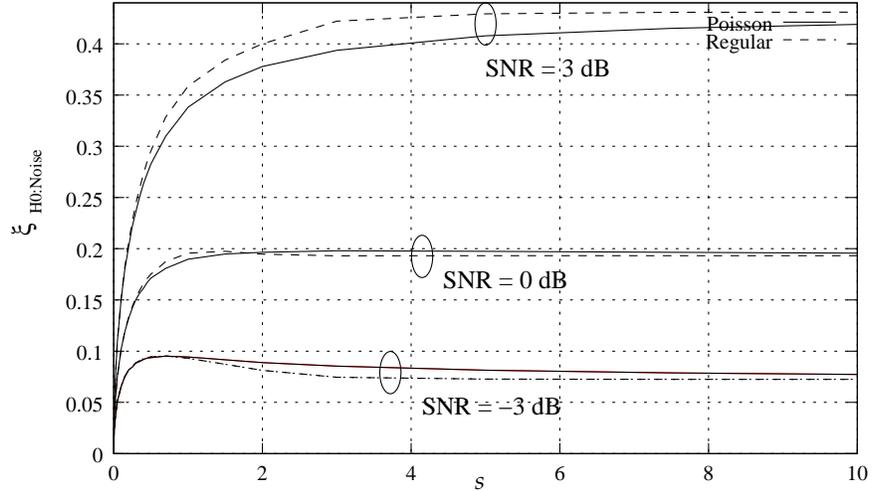}
  \end{center}
  \caption{RLC Model: $\xi_{\text{H0:Noise}}$ vs $s$ for
           $\text{SNR}=-3, 0$ and $3$ dB}
  \label{fig:H0-Noise-vector}
\end{figure}

\begin{figure}[t]
  \begin{center}
    \includegraphics[width=0.7\linewidth]{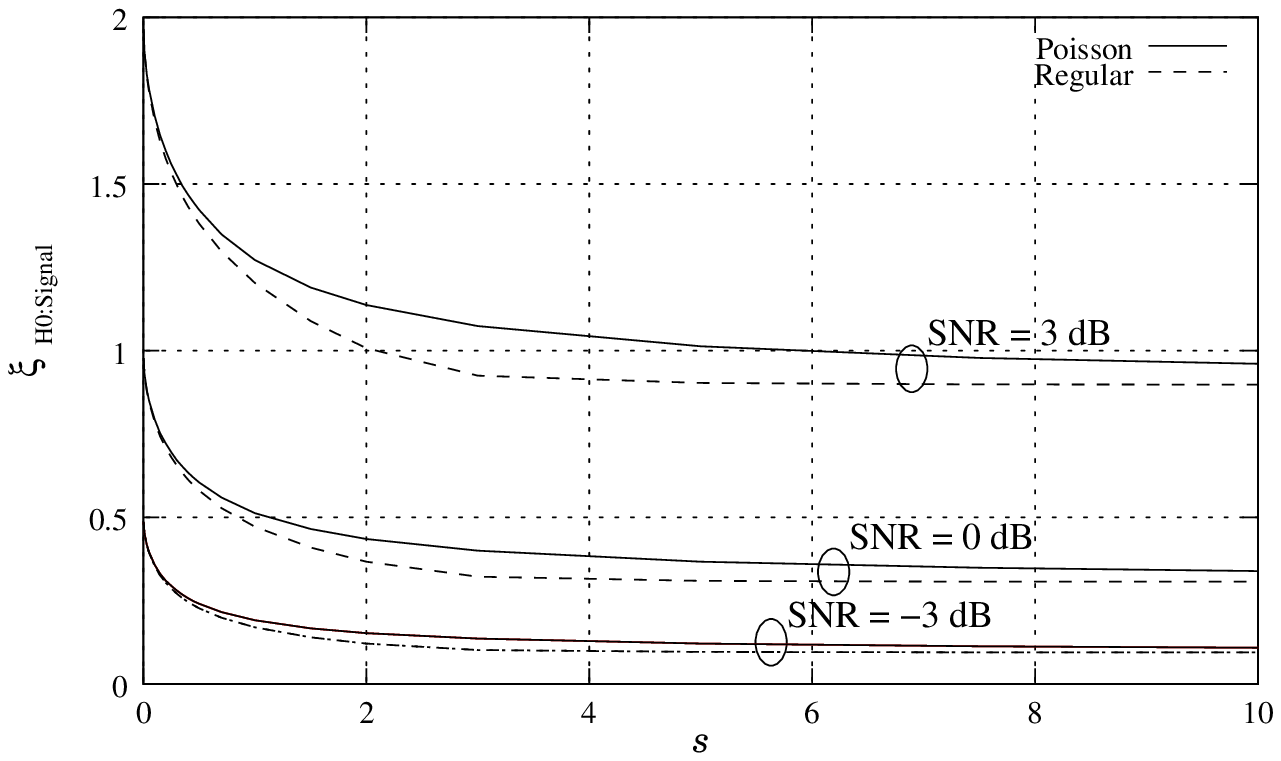}
  \end{center}
  \caption{RLC Model: $\xi_{\text{H0:Signal}}$ vs $s$ for
           $\text{SNR}=-3, 0$ and $3$ dB}
  \label{fig:H0-Signal-vector}
\end{figure}

Figures \ref{fig:H0-Noise-scalar} and \ref{fig:H0-Signal-scalar}
describe the behavior of the error exponents in the scalar case.
In Fig.~\ref{fig:H0-Noise-scalar}, $\xi_{\text{H0:Noise}}$ is plotted
as a function of $a$ for $\text{SNR} = -3, 0$ and $3$ dB.
Poisson sampling as well as regular sampling is considered in this figure.
We notice that $\xi_{\text{H0:Noise}}$
increases for $\text{SNR} =0$ and $3$ dB while it has a maximum with respect
to $a$ for $\text{SNR} = -3$ dB. As said in Section \ref{sec-discussion},
this behavior has been established in
\cite{sun-ton-poo-it06} in the case of a regular sampling. We also
notice that Poisson sampling is worse than regular sampling for
$\text{SNR} = 3$ dB and better than regular sampling for $\text{SNR} = -3$ dB
from the viewpoint of the error exponent. \\
In Fig.~\ref{fig:H0-Signal-scalar}, the error exponent
$\xi_{\text{H0:Signal}}$ is plotted \emph{vs} $a$ also for
$\text{SNR} = -3, 0$ and $3$ dB. The conclusions of Proposition
\ref{prop-H0:signal-errexp-monotone} are illustrated.
One interesting observation is that
the error exponent with Poisson sampling is better
than the error exponent with regular sampling for all considered $\text{SNR}$.
\\
Figures \ref{fig:H0-Noise-vector} and \ref{fig:H0-Signal-vector}
concern respectively the behavior of $\xi_{\text{H0:Noise}}$ and
$\xi_{\text{H0:Signal}}$ in the vector case.
We consider following 2-dimensional process.
$$
\rmd X(t) = - \begin{bmatrix} 0 & - 1 \\ 1 & 1 \end{bmatrix}
X(t) \, \rmd t +
\begin{bmatrix} 0 \\ 1 \end{bmatrix} \rmd W(t)
$$
where $W(t)$ is a scalar Brownian motion. We take $C=1_2$ in~(\ref{eq-H0:noise-vector}).
Both Poisson and regular models for the sampling are considered. In the
Poisson sampling case, we assume that the $I_n$ are equal in distribution
to $s I$ where $I$ is a Poisson random variable with mean one, and we plot
the error exponents in terms of the mean holding time $s$. In the regular
sampling case, $s$ is simply the sensor spacing. The other parameter is
the $\text{SNR}$ given by
\[
\text{SNR} =
\frac{\EE [| C X_n + V_n |^2]}{\EE[| V_n |^2]} =
\frac{\tr\left(C Q(\infty) C^\T \right)}{d} \ .
\]
A behavior comparable to the scalar case behavior is observed for both tests:
In the case of the H0-Noise test, the error exponent increases with $s$ at
high SNR, while it
has a maximum with respect to $s$ at low SNR, and the Poisson sampling is
worse than the regular sampling at high SNR. In the case of the H0-Signal
test, we also observe that $\xi_{\text{H0:Signal}}$ decreases in $s$,
and Poisson sampling is better than the regular sampling for the three
considered $\text{SNR}$ from the standpoint of the error exponents.

\appendix

\subsection{Technical lemmas}
\label{anx-technical}

In this section we provide some useful technical lemmas.
\begin{lemma}
\label{lm-norm-A}
Assume that $A$ is positive stable.
Then there exists constants $a > 0$ and $K > 0 $
such that $\| \rme^{-xA} \| \leq K \exp(-x a)$ for $x\geq 0$.
\end{lemma}
\begin{proof}
Let $a>0$ be smaller than the real parts of all the eigenvalues of $A$ and
$\mathcal{C}$ be a rectangle in the complex half plane $\{z:\Re(z)\geq a\}$ whose interior contains all these eigenvalues.
Applying Theorem~6.2.28 of \cite{hor-joh-topics-91}, we have
$\rme^{-xA}=\frac1{2\pi\,\rmi}\int_{\mathcal{C}}\rme^{-x\lambda}(\lambda \, I-A)^{-1}\,\rmd\lambda$.
Hence, we have
$$
\|\rme^{-xA}\|\leq \rme^{-x a} \int_{\mathcal{C}}\|(\lambda \, I-A)^{-1}\|\,\rmd\lambda.
$$
By continuity of $\lambda\mapsto(\lambda \, I-A)^{-1}$ on $\mathcal{C}$, the previous integral is finite, which gives the result.
\end{proof}

\begin{lemma}
\label{lm-properties-Q}
Assume that the matrix $A$ is positive stable and that the pair $(A,B)$ is controllable.
Then the matrix function $Q(x)$ defined by~(\ref{eq-def-Q(x)}) is strictly increasing in the positive semidefinite ordering
from $0$ to $Q(\infty)$ as $x$ increases from $0$ to $\infty$.
\end{lemma}
\begin{proof}
Since $(A,B)$ is controllable, $Q(x) > 0$ for any $x > 0$.
Assume that $x < y$. We have
$Q(y) - Q(x) = \int_x^y \exp({- u A}) B B^\T \exp({- u A^\T}) du
= \exp({-xA})  Q(y-x) \exp({-xA^\T}) > 0
$
which proves the lemma.
\end{proof}

\subsection{A stability result on Markov chains}
\label{anx-markov}

Here we present our swiss knife result on Markov chains.
We follow the approach in \cite{benveniste:metivier:priouret:1990} for obtaining the geometric ergodicity
of Markov chains using simple moment conditions, although we use a more direct proof inspired from~\cite{dia-fre-siam99}.
For the a.s. convergence of the empirical mean, we will rely on the following
standard result for martingales \cite{cho-65}:
\begin{lemma}
  \label{lem:AScvMart}
Let $(M_n)_{n\geq0}$ be a martingale sequence and $X_n=M_n-M_{n-1}$ be its increments. If there exists $p\in[1,2]$ such that
$\sum_{k\geq1}k^{-p}\EE[|X_n|^p]<\infty$, then $M_n/n\xrightarrow[n\to\infty]{\text{a.s.}} 0$.
\end{lemma}
We adopt the following setting for our generic Markov chain.
Let $\{\eta,\eta_k,\,k\geq1\}$ be an i.i.d. sequence of random variables valued in $\Eset$ and let $\Xset$ be a closed
subset of $\RR^d$.
Let $F_y(x)$ be defined for all $y\in \Eset$ and $x\in\Xset$ with values in $\Xset$ and such that $(x,y)\mapsto F_y(x)$ is
a measurable $\Xset\times\Eset\to\Xset$ function. This allows to define a Markov chain $\{Z_k^x,\,k\geq0\}$ by
\begin{equation}\label{eq:MCalaFD}
\left\{
\begin{split}
  &Z_0^x=x\;,\\
  &Z_{k}^x=F_{\eta_{k}}(Z_{k-1}^x),\quad k\geq1 \;.
\end{split}
\right.\end{equation}
This Markov chain is valued in $\Xset$ and start at time $0$ with the value $x$. We denote by $P$ the corresponding kernel
defined on any bounded continuous function $f:\Xset\to\RR$  by
$$
Pf (x) = \EE(f(Z_1^x))  = \EE(f\circ F_{\eta}(x)),\quad x\in\Xset\;.
$$
Observe that~(\ref{eq:MCalaFD}) implies, for
all $n\geq1$
$$
Z_{n}^x=F_{\eta_n}\circ\dots\circ F_{\eta_1}(x)\;.
$$

We denote by $|\cdot|$ the Euclidean norm on $\RR^d$ and, for any $p\geq1$ and $f:\Xset\to\RR$,
$$
\|f\|_{\lip_p}=\sup_{x,x'\in\Xset^2}\frac{|f(x)-f(x')|}{|x-x'|\,(1+|x|^{p-1}+|x'|^{p-1})} \; ,
$$
which is the Lipschitz norm for $p=1$.
We now state the main result of this appendix.

\begin{theorem}\label{thm:basicMarkov}
Define $\{Z_k^x,\,k\geq0\}$ as in~(\ref{eq:MCalaFD}).
Assume that $F_\eta$ is a.s. continuous, and that, for some $C>0$, $\alpha\in(0,1)$, $p,r\geq0$, $q\geq1$ and $s\geq p$,
\begin{enumerate}[(i)]
\item For all $x,x'\in\Xset^2$ and $n\geq1$,
$$
\EE\left[|F_{\eta_n}\circ\dots\circ F_{\eta_1}(x)-F_{\eta_n}\circ\dots\circ F_{\eta_1}(x')|^q\right]\leq C\alpha^n\,(1+|x|^{pq}+|x'|^{pq})\;.
$$
\item  For all $x\in\Xset^2$ and $n\geq1$,
  \begin{equation}
    \label{eq:unifMomentboundMC}
\EE\left[\left|F_{\eta_n}\circ\dots\circ F_{\eta_1}(x)\right|^s\right] \leq C (1+|x|^{rs}) \; .
  \end{equation}
\end{enumerate}
Then the following conclusions hold.
\begin{enumerate}[(a)]
\item There exists a unique probability measure $\mu$ on $\Xset$ such that
  \begin{equation}
    \label{eq:invMeas}
\xi\sim\mu\text{ and $\xi$ independent of $\eta$}\Rightarrow F_{\eta}(\xi)\sim \mu \;.
  \end{equation}
and this measure $\mu$ has a finite $s$-th moment.
\item Let $a\in[1, s\wedge\{1+s(q-1)/q\}]$  and  $f:\Xset\to\RR$ such that
  $\|f\|_{\lip_a}<\infty$. Suppose in addition that $s>b=p\vee \{r(a-1)\}$. Then, for all $x\in\Xset$,
$$
\frac1n \sum_{k=1}^n f(Z_k^x)\xrightarrow[n\to\infty]{\text{a.s.}}  \mu(f) =
\int f d\mu \;.
$$
\item Let $(U_n)_{n\geq1}$ be a sequence of i.i.d. real-valued random variables such that $\EE[|U_1|^{1+\epsilon}]<\infty$ for some
  $\epsilon>0$ and, for all $n\geq1$, $U_n$ is independent of $\eta_1,\dots,\eta_n$. Then, under the same assumptions as in
  (b), if moreover $s>a$, then, for all $x\in\Xset$,
$$
\frac1n \sum_{k=1}^n U_kf(Z_k^x)\xrightarrow[n\to\infty]{\text{a.s.}}  m\;\mu(f)\;,
$$
where $m=\EE[U_1]$.
\end{enumerate}
\end{theorem}
\begin{proof}
Let us introduce the backward recurrence process starting at $x$ defined by $Y_0=x$ and
$$
Y_n=F_{\eta_1}\circ\dots\circ F_{\eta_n}(x),\quad n\geq1\;.
$$
Note that for any $n$, $Y_n\stackrel{d}{=}Z_n^{x}$, that is, the processes
$(Y_n)$ and $(Z_n^x)$ has the same marginal
distributions. Moreover, using (i) and the Jensen Inequality, we have
$$
\EE\left[\sum_{n\geq0}\left|Y_{n+1}-Y_n\right|\right] \leq \sum_{n\geq0} C^{1/q}\alpha^{n/q} \;
(1+|x|^{p}+\EE\left[|F_{\eta_n}(x)|^{p}\right])\;.
$$
By (ii), since $s\geq p$, $\EE\left[|F_{\eta_n}(x)|^{p}\right]<\infty$ and thus $\sum_{n\geq0}\left|Y_{n+1}-Y_n\right|<\infty$ a.s.
By completeness of the state space $\Xset$, $Y_n$ converges in $\Xset$ a.s. We denote the limit by $\xi$ and its
probability distribution by $\mu$. By a.s. continuity of $F_\eta$, we have
$F_\eta(Y_n)\stackrel{a.s.}{\to}F_\eta(\xi)$.
On the other hand $F_\eta(Y_n)\sim Y_{n+1}\stackrel{a.s.}{\to}\xi$. Hence $\mu$
satisfies~(\ref{eq:invMeas}), that is, $\mu$ is an invariant
distribution of the induced Markov chain. Moreover by (ii), we have
$\sup_n\EE[|Y_n|^s]<\infty$ which by Fatou's Lemma implies that $\EE[|\xi|^s]<\infty$.
Let us show that $\mu$ is the unique invariant distribution. By (i), for any $x,y\in\Xset$,
$Z_n^x -Z_n^y\xrightarrow[n\to\infty]{\text{a.s.}} 0$.
Now draw $x$ and $y$ according to two invariant distributions, respectively, so that $(Z_n^x)_{n\geq0}$ and
$(Z_n^y)_{n\geq0}$ are two sequences with constant marginal distributions. Then necessary these two distributions are the
same and thus $\mu$ is the unique invariant distribution, which achieves the proof of (a).

We now prove (b). First observe that $f$ is continuous and $f(x)=O(|x|^a)$ as $|x|\to\infty$.
Hence by~(a), since $a\leq s$, $f$ is integrable with respect to $\mu$. Also, by (ii), $\EE[|f(Z_k^x)|]<\infty$
for all $k\geq1$ and $x\in\Xset$.
We use the classical Poisson equation for decomposing the empirical mean of the Markov chain
as the empirical mean of martingale increments plus a negligible remainder.
Using that $\|f\|_{\lip_a}<\infty$, the Hölder inequality and (i), we have, for any $x,y\in\Xset$,
$$
\EE\sum_{k\geq1}|f(Z_k^x)-f(Z_k^y)|\leq \sum_{k\geq1}C^{1/q}\alpha^{k/q}\,
(1+\|Z_k^x\|_{q'(a-1)}^{a-1}+\| Z_k^y\|_{q'(a-1)}^{a-1})\;(1+|x|^{p}+|y|^{p})\;,
$$
where we used the notation $\|\cdot\|_{p}=(\EE[|\cdot|^p])^{1/p}$ and $q'=q/(q-1)$. Since $q'(a-1)\leq s$, we can apply the
Jensen Inequality and~(ii) to bound $\|Z_k^x\|_{q'(a-1)}$ and $\|Z_k^y\|_{q'(a-1)}$.
We obtain, for some constant $c>0$
$$
\EE\sum_{k\geq1}|f(Z_k^x)-f(Z_k^y)|\leq c \; (1+|x|^{b}+|y|^{b})\;.
$$
with $b=p\vee \{r(a-1)\}$.
Since we assumed $s>b$, using (a), the right-hand side of the previous display is integrable in $y$ with respect
to $\mu$ and we get
$$
\sum_{k\geq1}|\EE[f(Z_k^x)]-\mu(f)|\leq \int\EE\sum_{k\geq1}|f(Z_k^x)-f(Z_k^y)|\,\mu(\rmd y)\leq c'\,(1+|x|^{b})\;.
$$
Hence we may define the real-valued function
$$
\hat{f}(x)=\sum_{k\geq1}\{\EE[f(Z_k^x)]-\mu(f)\}\;,
$$
which is the solution of the Poisson equation $f(x)-\mu(f)=\hat{f}(x)-P\hat{f}(x)$
and satisfies
\begin{equation}
  \label{eq:momentbounforpoisson}
\sup_{x\in\Xset}(1+|x|^{b})^{-1}\sup_{k\geq1}|\hat{f}(x)|<\infty \; .
\end{equation}
Hence the decomposition
$$
\frac1n \sum_{k=1}^n \{f(Z_k^x)- \mu(f)\}=\frac1n \sum_{k=1}^n \{\hat{f}(Z_k^x)-P\hat{f}(Z_{k}^x)\}
=\frac1n \sum_{k=1}^n X_k+\frac1n\{P\hat{f}(x) - P\hat{f}(Z_n^x)\}\;,
$$
where $X_k=\hat{f}(Z_k^x)-P\hat{f}(Z_{k-1}^x)$, $k\geq1$. Observe that $(X_k)_{k\geq1}$ is a sequence of martingale
increments. By the Jensen Inequality, we have $\EE[|P\hat{f}(Z_{n}^x)|^{s/b}]\leq\EE[|\hat{f}(Z_{n+1}^x)|^{s/b}]$
and by~(\ref{eq:momentbounforpoisson}) and (ii), $\sup_{k\geq1}\EE[|\hat{f}(Z_{n+1}^x)|^{s/b}]<\infty$.
Since $s/b>1$, by the Markov Inequality and Borel-Cantelli's lemma, this implies that
$P\hat{f}(Z_n^x)/n\stackrel{a.s.}{\to}0$. We also get that $\sup_{k\geq1}\EE[|X_k|^{s/b}]<\infty$
and, by Lemma~\ref{lem:AScvMart} $\sum_{k=1}^n X_k/n\stackrel{a.s.}{\to}0$. This proves (b).

We conclude with the proof of (c). Using (b) we may replace $U_k$ by $U_k-m$, that is, we assume $m=0$ without loss of
generality. Then $(U_kf(Z_k^x))_{k\geq1}$ is a sequence of martingale increments. Let $u=(1+\epsilon)\wedge s/a>1$.
We have $\sup_{k\geq1}\EE[|U_kf(Z_k^x)|^u]=\EE[|U_1|^u]\sup_{k\geq1}\EE[|f(Z_k^x)|^u]<\infty$ by (ii) and the result follows
from Lemma~\ref{lem:AScvMart}.
\end{proof}

\subsection{Proofs for Section \ref{sec-discussion}.}
\label{anx-proofs-particular}

\subsubsection{Proof of Proposition \ref{prop-regular-sampling}}
Given any deterministic nonnegative matrix ${\bf p} \in [0, Q(\infty)]$, the
sequence of covariance matrices $\tilde{Z}^{\bf p}_k =
\tilde{F}_1(\tilde{Z}^{\bf p}_{k-1})$ where $\tilde F_1$ is the second
component of \eqref{eq:PsiEta} with $I=1$ is a deterministic sequence. From
Lemma \ref{lm-decrease-upsilon} and
Proposition \ref{prop:main-Lipschitz-bounds}--Eq. \eqref{eq:LipschitzP1},
we have 
$\| \tilde{Z}^{\bf p}_{k} -\tilde{Z}^{\bf q}_{k} \| \leq
K \alpha^k \| {\bf p} - {\bf q} \|$
for $\alpha \in (0,1)$ and $K > 0$. Hence, (see the classical proof of 
the Banach fixed point theorem),  
$\tilde{Z}^{\bf p}_{k}$ converges to a limit $P_{\text{R}}$ defined as the
unique solution in $[0, Q(\infty)]$ of the equation $P = \tilde F_1(P)$ which
is the discrete algebraic Riccati equation \eqref{eq-riccati}.
In the formalism of Proposition \ref{prop-invariance-vector-P}, this amounts
to say that the invariant distribution $\mu$ coincides with
$\delta_{P_{\text{R}}}$. It remains to show that Equation \eqref{eq-riccati}
has no solutions outside $[0, Q(\infty)]$. Indeed, assume that ${\bf p}$
is a solution of \eqref{eq-riccati}. Consider the state
equations \eqref{eq-Y:signal-vector} where it is assumed that
$X(0) \sim {\cal N}(0, {\bf p})$. By the very nature of the Kalman filter,
the covariance matrix $P_k$ satisfies
\[
P_k \leq \EE\left[ X_k X_k^\T \right]
= \rme^{-T_k A} {\bf p} \rme^{-T_k A^\T} + Q(T_k)
< \rme^{-T_k A} {\bf p} \rme^{-T_k A^\T} + Q(\infty)
\]
As $P_k = {\bf p}$ for any $k$, we have
${\bf p} \leq Q(\infty)$ by taking the limit as $k\to\infty$. \\
We now consider the invariant distribution $\nu$ characterized by Proposition
\ref{prop-invariance-vector-W}.
This distribution writes $\nu = \nu_X \otimes \delta_{P_{\text{R}}}$, and
we shall show that $\nu_X = {\cal N}(0, \Sigma)$ where $\Sigma$ is the
unique solution of Equation \eqref{eq-Sigma-regular-sampling}.
To that end, we begin by showing that the steady state Kalman filter transition
matrix $\Theta = \Phi( 1_q - G C )$ with $G = P_{\text{R}} C^\T (
C P_{\text{R}} C^\T + 1_d)^{-1}$ has all its eigenvalues $\{ \lambda_i \}$
in the open unit disk. Indeed, getting back to Equation
\eqref{eq-update-P-transition-matrix} and passing to the limit, we have
$P_{\text{R}} = \Theta P_{\text{R}} \Theta^\T + \Phi G G^\T \Phi^\T + Q$.
Assuming $t_i$ is an eigenvector of $\Theta$ with eigenvalue $\lambda_i$, we
obtain from this last equation that
$(1 - | \lambda_i |^2) t_i^\T P_{\text{R}} t_i =
t_i^\T \Phi G G^\T \Phi^\T t_i + t_i^\T Q t_i > 0
$
due to $Q = Q(1) > 0$, hence $| \lambda_i | < 1$.  Consequently, the matrix
equation \eqref{eq-Sigma-regular-sampling} has a unique solution
$\Sigma = \sum_{n=0}^\infty \Theta^n \Phi G G^\T \Phi^\T (\Theta^\T)^n$
\cite[Chap.~4.2]{and-moo-livre79}. When
$Z_k = (\underline{Z}_k, \tilde Z_k) \in \RR^q \times [0, Q(\infty)]$
follows the distribution $\nu$, we have (see \eqref{eq:PsiEta})
$\underline{Z}_k = \Theta \underline{Z}_{k-1} + \Phi G Y_{k}$. Recall that
$Y_{k} \sim {\cal N}(0, 1_d)$ and is independent with $\underline{Z}_{k-1}$.
In these
conditions, it is clear that $\underline{Z}_k \sim {\cal N}(0, \Sigma)$ when
$\underline{Z}_{k-1} \sim {\cal N}(0, \Sigma)$. Therefore,
$\nu = {\cal N}(0, \Sigma) \otimes \delta_{P_{\text{R}}}$ is invariant, and
by Proposition \ref{prop-invariance-vector-W}, it is the unique invariant
distribution. Replacing $\nu$ and $\mu$ with their values at the right hand
sides of \eqref{eq-err-exp-H0:bruit-vector} and
\eqref{eq-err-exp-H0:signal-vector}, we obtain
\eqref{eq-exposant-H0:Noise-regular} and \eqref{eq-exposant-H0:Signal-regular}
respectively. Proposition \ref{prop-regular-sampling} is proven.

\subsubsection{Proof of Proposition \ref{prop-large-s}}
We assume that the holding times $I_n$ are equal in distribution to $I^s$
(distributed as $\tau_s$) to point out the dependence on $s$.
We also denote the invariant distribution of the Markov chain $(\tilde Z_k)$
defined in Section \ref{sec-proofs-vector} as $\mu_s$. We begin by proving
that $\mu_s$ converges weakly to
$\delta_{Q(\infty)}$ as $s\to\infty$ (notation $\mu_s \Rightarrow
\delta_{Q(\infty)}$). By Lemma \ref{lm-norm-A} we have
$\EE [ \| \exp(-I^{s} A ) \|^2 ] \leq K \EE [ \exp(- 2 a I^{s} ) ] =
\int \exp(-2ax) \tau_s(\rmd x)$ with $a > 0$. Given a $K > 0$, we have
$\int \exp(-2ax) \tau_s(\rmd x) = \int_0^K \exp(-2ax) \tau_s(\rmd x) +
\int_K^\infty \exp(-2ax) \tau_s(\rmd x) \leq \tau_s([0,K]) + \exp(-2aK)$.
As $\tau_s$ escapes to infinity, $\EE[ \| \rme^{-I^{s}A} \|^2]
\rightarrow_{s\to\infty} 0$, which implies that $\rme^{-I^{s}A}
\to_{s\to\infty} 0$ in probability. Moreover, we have
\begin{multline*}
\| Q(I^{s}) - Q(\infty) \| =
\left\| \int_{I^{s}}^\infty \exp(-u A) B B^\T \exp(-u A^\T) \, \rmd u \right\|
\\
\leq
\| B \|^2 \int_{I^{s}}^\infty \| \exp(-u A) \|^2 \, \rmd u
\leq
K \int_{I^{s}}^\infty \exp(-2 u a) \, \rmd u
= (K/2a) \exp(-2a I^{s})
\end{multline*}
hence
$Q(I^{s}) \to Q(\infty)$ in probability as $s\to\infty$. Now, assume that the
random variable $\tilde{Z} \in [0, Q(\infty)]$ is distributed as $\mu_s$.
Recalling that $\tilde F$ is the
random iteration function defined as the second component of Equation
\eqref{eq:PsiEta}, we have
$\| \tilde{F}_{I^s}(\tilde{Z}) - Q(\infty) \| \leq K \| \rme^{-I^{s}A} \|^2 +
\| Q_{I^s} - Q(\infty) \|$, hence $\tilde{F}_{I^s}(\tilde{Z}) \to Q(\infty)$
in probability as $s\to\infty$. As $\tilde{F}_{I^s}(\tilde{Z}) \sim \mu_s$,
$\mu_s \Rightarrow \delta_{Q(\infty)}$. Due to the continuity of the $\log\det$
on the compact set $[ 0, Q(\infty) ]$, we have
$\int \log(1+{\bf p}) \rmd \mu_s({\bf p}) \rightarrow_{s\to\infty} \int
\log(C Q(\infty) C^\T + 1_d)$, and \eqref{eq-xi-H0:Signal-large-s} results from
\eqref{eq-err-exp-H0:signal-vector}. \\
Now assume that $Z=(\underline{Z}, \tilde{Z} ) \in \RR^q \times [0, Q(\infty)]$
follows the invariant distribution $\nu$, and let
$(\underline{Z}_1, \tilde{Z}_1) = F_{(I^s, V)}(Z)$, where $F_{(I^s,V)}$ is
defined by Equation \eqref{eq:PsiEta}. In particular, we have
$\underline{Z}_1 = \Theta(I^s, \tilde{Z}) \underline{Z} + \rme^{-I^{s} A}
G(\tilde{Z}) V$. As $\EE[ \| \rme^{-I^{s}A} \|^2]
\rightarrow_{s\to\infty} 0$ and $\tilde{Z} \leq Q(\infty)$, we have
$\EE[ \| \Theta(I^s, \tilde{Z}) \|^2 ] =
\EE[ \| \rme^{-I^{s}A} (I - G(\tilde{Z}) C) \|^2 ]
\rightarrow_{s\to\infty} 0$ and $\EE[ | \rme^{-I^{s}A} G(\tilde{Z}) V |^2 ]
\rightarrow_{s\to\infty} 0$, hence
$\EE[ | \underline{Z}_1 |^2] \rightarrow_{s\to\infty} 0$.
The third term at the RHS of the Expression
\eqref{eq-err-exp-H0:bruit-vector} of $\xi_{\text{H0:Noise}}$ satisfies
\[
\int {\bf x}^\T C^\T \left( C {\bf p} C^\T + 1_d \right)^{-1} C {\bf x} \
\rmd \nu({\bf x}, {\bf p})
\leq \| C \|^2 \int | {\bf x} | ^2 \rmd \nu({\bf x}, {\bf p}) =
\| C \|^2 \EE[ | \underline{Z}_1 |^2] \xrightarrow[s\to\infty]{}  0 \ .
\]
As $\mu_s \Rightarrow \delta_{Q(\infty)}$, the second term at the RHS of
\eqref{eq-err-exp-H0:bruit-vector} converges to
$- \tr[ C Q(\infty) C^\T (C Q(\infty) C^\T + 1_d)^{-1} ]$, which terminates
the proof of Proposition \ref{prop-large-s}.

\subsubsection{Proof of Proposition \ref{prop-H0:signal-errexp-monotone}}
In the scalar case, the covariance update equation \eqref{eq-kalman-vector-P}
writes
\begin{equation}
\label{eq-recursion-kalman-P-scalar}
P_{n+1} = \tilde{F}^{a}_{I_{n+1}}(P_n) = \rme^{-2 a I_{n+1}}
\left( \frac{P_n}{ P_n  + 1} - Q(\infty) \right)
+ Q(\infty) \ .
\end{equation}
Given a sequence of holding times $(I_n)_{n\geq 1}$ and two positive numbers
$a_1 \geq a_2$, consider the two Markov chains
$\tilde{Z}^{p}_{a_i, k} =
\tilde{F}^{a_i}_{I_{k}}( \tilde{Z}^{p}_{a_i, k-1})$ for $i=1,2$, both starting
at the same value ${p} = Q(\infty)$. Let $f(p) = p / (p+1) - Q(\infty)$.
As $f(Q(\infty)) < 0$ and $0 < \exp(-2 a_1 I_1) \leq \exp(-2 a_2 I_1)$,
it is clear that $\tilde{Z}^{p}_{a_1, 1} \geq \tilde{Z}^{p}_{a_2, 1}$.
Assume that $\tilde{Z}^{p}_{a_1, k-1} \geq \tilde{Z}^{p}_{a_2, k-1}$.
As $f(p)$ is negative and increasing for $p \in [0,Q(\infty)]$ and
$0 < \exp(-2 a_1 I_{k}) \leq \exp(-2 a_2 I_{k})$, we have
$\tilde{Z}^{p}_{a_1, k} =
\exp(-2 a_1 I_{k}) f(\tilde{Z}^{p}_{a_1, k-1}) + Q(\infty) \geq
\exp(-2 a_2 I_{k}) f(\tilde{Z}^{p}_{a_2, k-1}) + Q(\infty) =
\tilde{Z}^{p}_{a_2, k}$. \\
From Proposition \ref{prop-invariance-vector-P}, both the chains
$\tilde{Z}^{p}_{a_1, k}$ and $\tilde{Z}^{p}_{a_2, k}$ have unique
invariant distributions $\mu_1$ and $\mu_2$ respectively, and by repeating
the arguments of the proof of Theorem \ref{th-H0:Signal-vector},
\[
\frac{1}{N} \sum_{k=0}^{N-1} \log\left( 1 + \tilde{Z}^{p}_{a_i, k}\right)
\xrightarrow[N\to\infty]{\text{a.s.}}
\int \log\left( 1 + p \right) \rmd \mu_i(p)
\quad \text{for} \  i=1,2.
\]
As $\tilde{Z}^{p}_{a_1, k} \geq \tilde{Z}^{p}_{a_2, k}$ for all $k$,
by passing to the limit we have $\int \log( 1 + p ) \rmd \mu_1(p)
\geq \int \log( 1 + p) \rmd \mu_2(p)$.
As $\xi_{\text{H0:Signal}} = 0.5 \left( Q(\infty) -
\int \log( 1 + p ) \rmd \mu(p) \right)$ in the scalar case (see Expression
\eqref{eq-err-exp-H0:signal-vector}), this error exponent decreases with $a$.\\
We now show that $\lim_{a \to 0} \xi_{\text{H0:Signal}} = Q(\infty) / 2$.
Assume that $\tilde{Z} \in [0, Q(\infty)]$ has the invariant distribution
that we denote $\mu_a$. From Eq. \eqref{eq-recursion-kalman-P-scalar}, we have
$\EE[\tilde{Z}] = \EE[ \tilde{F}^{a}_{I}(\tilde{Z}) ] =
\EE[\rme^{-2 a I}] \left( \EE\left[\frac{\tilde{Z}}{\tilde{Z}+1} - Q(\infty)
\right] \right) + Q(\infty)$
which results in
$$
\EE\left[ \frac{\tilde{Z}^2 + (1 - \EE[\rme^{-2 a I}]) \tilde{Z}}
{\tilde{Z} + 1} \right] = Q(\infty) (1 - \EE[\rme^{-2 a I}]) \ .
$$
As $\tilde{Z} \leq Q(\infty)$, we have
$
\EE\left[ \frac{\tilde{Z}^2}{Q(\infty)+1} \right] \leq
\EE\left[ \frac{\tilde{Z}^2}{\tilde{Z}+1} \right] \leq
Q(\infty)(1 - \EE[\rme^{-2 a I}])
$.
By the dominated convergence theorem, $\EE[ \exp(-2a I)] \to_{a\to 0} 1$,
therefore $\EE[\tilde{Z}^2] \to 0$ as $a\to 0$.
It results that $\mu_a$ converges weakly to $\delta_0$ as $a\to 0$, therefore
$\int \log(1+p) \rmd \mu_a(p) \to 0$. Hence
$\lim_{a \to 0} \xi_{\text{H0:Signal}} = Q(\infty) / 2$. \\
In order to show that $\xi_{\text{H0:Signal}}$ increases with $Q(\infty)$,
the argument
is similar to the one used above to show that $\xi_{\text{H0:Signal}}$
decreases as $a$ increases.
Proposition \ref{prop-H0:signal-errexp-monotone} is proven.

\bibliographystyle{IEEEbib}

\begin{thebibliography}{10}

\bibitem{kai-poo-it98}
T.~Kailath and H.V. Poor,
\newblock ``Detection of stochastic processes,''
\newblock {\em IEEE Trans. on Information Theory}, vol. 44, no. 6, pp.
  2230--2259, Oct. 1998.

\bibitem{mis-ton-sp08}
S.~Misra and L.~Tong,
\newblock ``Error exponents for the detection of {G}auss-{M}arkov signals using
  randomly spaced sensors,''
\newblock {\em IEEE Trans. on Signal Processing}, vol. 56, no. 8, pp.
  3385--3396, Aug. 2008.

\bibitem{oks-livre03}
B.~\O ksendal,
\newblock {\em {Stochastic Differential Equations: An Introduction With
  Applications}},
\newblock Springer Verlag, 6th edition, 2003.

\bibitem{all-livre2007}
E.~Allen,
\newblock {\em {Modeling with It\^{o} Stochastic Differential Equations}},
  vol.~22 of {\em Mathematical Modelling: Theory and Applications},
\newblock Springer, 2007.

\bibitem{mic-01}
M.~Micheli,
\newblock ``{Random Sampling of a Continuous-Time Stochastic Dynamical System:
  analysis, State Estimation, and Applications},''
\newblock M.S. thesis, UC Berkeley, 2001.

\bibitem{mic-jor-mtns02}
M.~Micheli and M.I. Jordan,
\newblock ``Random sampling of a continuous-time stochastic dynamical system,''
\newblock in {\em Proc. of the International Symposium on the Mathematical
  Theory of Networks and Systems}, Univ. of Notre Dame, South Bend, Indiana,
  USA, Aug. 2002.

\bibitem{sun-ton-poo-it06}
Y.~Sung, L.~Tong, and H.V. Poor,
\newblock ``{N}eyman-{P}earson detection of {G}auss-{M}arkov signals in noise:
  Closed-form error exponents and properties,''
\newblock {\em IEEE Trans. on Information Theory}, vol. 52, no. 4, pp.
  1354--1365, Apr. 2006.

\bibitem{cha-vee-sp03}
J.-F. Chamberland and V.V. Veeravalli,
\newblock ``Decentralized detection in sensor networks,''
\newblock {\em IEEE Trans. on Signal Processing}, vol. 51, no. 2, pp. 407--416,
  Feb. 2003.

\bibitem{cha-vee-it06}
J.-F. Chamberland and V.V. Veeravalli,
\newblock ``How dense should a sensor network be for detection with correlated
  observations ?,''
\newblock {\em IEEE Trans. on Information Theory}, vol. 52, no. 11, pp.
  5099--5106, Nov. 2006.

\bibitem{jay-sp07}
S.K. Jayaweera,
\newblock ``Bayesian fusion performance and system optimization for distributed
  stochastic {G}aussian signal detection under communication constraints,''
\newblock {\em IEEE Trans. on Signal Processing}, vol. 55, no. 4, pp.
  1238--1250, Apr. 2007.

\bibitem{sun-mis-ton-eph-spmag06}
Y.~Sung, S.~Misra, L.~Tong, and A.~Ephremides,
\newblock ``Signal processing for application-specific ad hoc networks,''
\newblock {\em IEEE Signal Processing Magazine}, vol. 23, no. 5, pp. 74--83,
  Sept. 2006.

\bibitem{sun-mis-ton-eph-jsac07}
Y.~Sung, S.~Misra, L.~Tong, and A.~Ephremides,
\newblock ``Cooperative routing for distributed detection in large sensor
  networks,''
\newblock {\em IEEE Journal on Selected Areas in Communications}, vol. 25, no.
  2, pp. 471--483, Feb. 2007.

\bibitem{che-it96}
Po-Ning Chen,
\newblock ``{General formulas for the Neyman-Pearson type-II error exponent
  subject to fixed and exponential type-I error bounds},''
\newblock {\em IEEE Trans. on Information Theory}, vol. 42, no. 1, pp.
  316--323, Jan 1996.

\bibitem{che-ams52}
H.~Chernoff,
\newblock ``A measure for asymptotic efficiency for tests of a hypothesis based
  on a sum of observations,''
\newblock {\em Ann. Math. Statist.}, vol. 23, no. 4, pp. 493--507, 1952.

\bibitem{lus-ruk-vaj-spa93}
H.~Luschgy, A.~Rukhin, and I.~Vajda,
\newblock ``Adaptive tests for stochastic processes in the ergodic case,''
\newblock {\em Stoch. Process. Appl.}, vol. 45, no. 1, pp. 45--59, 1993.

\bibitem{ana-ton-swa-icassp07}
A.~Anandkumar, L.~Tong, and A.~Swami,
\newblock ``Detection of {G}auss-{M}arkov random field on nearest-neighbor
  graph,''
\newblock in {\em IEEE International Conference on Acoustics, Speech, and
  Signal Processing}, Hawaii, USA, 2007.

\bibitem{sun-zha-ton-poo-sp08}
Y.~Sung, X.~Zhang, L.~Tong, and H.V. Poor,
\newblock ``Sensor configuration and activation for field detection in large
  sensor arrays,''
\newblock {\em IEEE Trans. on Signal Processing}, vol. 56, no. 2, pp. 447--463,
  Feb. 2008.

\bibitem{bha-it90}
R.K. Bahr,
\newblock ``Asymptotic analysis of error probabilities for the nonzero-mean
  {G}aussian hypothesis testing problem,''
\newblock {\em IEEE Trans. on Information Theory}, vol. 36, no. 3, pp.
  597--607, May 1990.

\bibitem{sch-it65}
F.~Schweppe,
\newblock ``Evaluation of likelihood functions for {G}aussian signals,''
\newblock {\em IEEE Trans. on Information Theory}, vol. 11, no. 1, pp. 61--70,
  Jan. 1965.

\bibitem{hor-joh-topics-91}
R.~Horn and C.~Johnson,
\newblock {\em {Topics in Matrix Analysis}},
\newblock Cambridge Univ. Press, 1991.

\bibitem{bro-dav-livre91}
P.~Brockwell and R.~Davis,
\newblock {\em {Time Series: Theory and Methods}},
\newblock Springer Series in Statistics, 1991.

\bibitem{ros-livre70}
H.H. Rosenbrock,
\newblock {\em State Space and Multivariable Theory},
\newblock Nelson, 1970.

\bibitem{kar-sin-mou-arxiv-09}
S.~Kar, B.~Sinopoli, and J.~Moura,
\newblock ``{Kalman Filtering with Intermittent Observations: Weak Convergence
  to a Stationary Distribution},''
\newblock {submitted to \emph{IEEE Trans. Automatic Control}, [online]
  \texttt{arXiv:0903.2890v1}}, Mar. 2009.

\bibitem{guo:1994}
L.~Guo,
\newblock ``Stability of recursive stochastic tracking algorithms,''
\newblock {\em SIAM J. on Control and Optimization}, vol. 32, pp. 1195--1125,
  1994.

\bibitem{and-moo-siam81}
B.D.O. Anderson and J.B. Moore,
\newblock ``Detectability and stabilizability of time-varying discrete-time
  linear systems,''
\newblock {\em SIAM J. Control and Optimization}, vol. 19, no. 1, pp. 20--32,
  1981.

\bibitem{and-moo-livre79}
B.D.O. Anderson and J.B. Moore,
\newblock {\em {Optimal Filtering}},
\newblock Prentice-Hall, 1979.

\bibitem{benveniste:metivier:priouret:1990}
A.~Benveniste, M.~M{\'e}tivier, and P.~Priouret,
\newblock {\em Adaptive algorithms and stochastic approximations}, vol.~22 of
  {\em Applications of Mathematics (New York)},
\newblock Springer-Verlag, Berlin, 1990,
\newblock Translated from the French by Stephen S. Wilson.

\bibitem{dia-fre-siam99}
P.~Diaconis and D.~Freedman,
\newblock ``Iterated random functions,''
\newblock {\em SIAM Review}, vol. 41, no. 1, pp. 45--76, Mar. 1999.

\bibitem{cho-65}
Y.S. Chow,
\newblock ``Local convergence of martingales and the law of large numbers,''
\newblock {\em Ann. Math. Statist.}, vol. 36, no. 2, pp. 552--558, 1965.

\end{thebibliography}

\def\cprime{$'$}

\end{document}